\documentclass[]{interact}

\usepackage{tabularx}

\usepackage{xcolor} % for manuscript highlighting purposes
\usepackage{mdframed} % for manuscript highlighting purposes

\usepackage{epstopdf}% To incorporate .eps illustrations using PDFLaTeX, etc.
\usepackage[caption=false]{subfig}% Support for small, `sub' figures and tables

\usepackage{url}

\usepackage[natbibapa,nodoi]{apacite}
\theoremstyle{plain}% Theorem-like structures provided by amsthm.sty

\theoremstyle{definition}

\theoremstyle{remark}

\begin{document}

%\articletype{ARTICLE TEMPLATE}% Specify the article type or omit as appropriate

This is an Accepted Manuscript of an article published by Taylor \& Francis in New Review of Hypermedia and Multimedia on February 9, 2021, available online: http://www.tandfonline.com/doi/abs/10.1080/13614568.2021.1889691.

\title{A Study of Fake News Reading and Annotating in Social Media Context}
% 
% Original title: Fake News Reading on Social Media: An Eye-tracking Study
%
% \color{Blue}
% \color{Black}

%\author{
%\name{A.~N. Author\textsuperscript{a}\thanks{CONTACT A.~N. Author. Email: latex.helpdesk@tandf.co.uk} and John Smith\textsuperscript{b}}
%\affil{\textsuperscript{a}Taylor \& Francis, 4 Park Square, Milton Park, Abingdon, UK; \textsuperscript{b}Institut f\"{u}r Informatik, Albert-Ludwigs-Universit\"{a}t, Freiburg, Germany}
%}

\author{
\name{Jakub Simko\textsuperscript{a}\thanks{CONTACT J. Simko. Email: jakub.simko@kinit.sk},  Patrik Racsko\textsuperscript{b}, Matus Tomlein\textsuperscript{a}, Martina Hanakova\textsuperscript{b}, Robert Moro\textsuperscript{a}, Maria Bielikova\textsuperscript{a}}
\affil{\textsuperscript{a}Kempelen Institute of Intelligent Technologies, Mlynske nivy 5, 821 09, Bratislava Slovakia\\
\textsuperscript{b}Slovak University of Technology in Bratislava, Ilkovicova 2, 842 16, Bratislava, Slovakia}
}

\maketitle

\begin{abstract}
The online spreading of fake news is a major issue threatening entire societies. Much of this spreading is enabled by new media formats, namely social networks and online media sites. Researchers and practitioners have been trying to answer this by characterizing the fake news and devising automated methods for detecting them. The detection methods had so far only limited success, mostly due to the complexity of the news content and context and lack of properly annotated datasets. One possible way to boost the efficiency of automated misinformation detection methods, is to imitate the detection work of humans. It is also important to understand the news consumption behavior of online users. In this paper, we present an eye-tracking study, in which we let 44 lay participants to casually read through a social media feed containing posts with news articles, some of which were fake. In a second run, we asked the participants to decide on the truthfulness of these articles. We also describe a follow-up qualitative study with a similar scenario but this time with 7 expert fake news annotators. We present the description of both studies, characteristics of the resulting dataset (which we hereby publish) and several findings.

%TODO primerane doplnit informaciu o follow-up studii

% The online spreading of fake news is a major issue threatening entire societies. Much of this spreading is enabled by new media formats, namely social networks and online media sites. Researchers and practitioners have been trying to answer this by characterizing the fake news and devising automated methods for detecting them. The detection methods had so far only limited success, mostly due to the complexity of the news content and context and lack of properly annotated datasets. One possible way to boost the efficiency of automated misinformation detection methods, is to imitate the detection work of humans. It is also important to understand the news consumption behavior of online users. In this paper, we present an eye-tracking study, in which we let 44 lay participants to casually read through a social media feed containing posts with news articles, some of which were fake. In a second run, we asked the participants to decide on the truthfulness of these articles. We also describe a follow-up qualitative study with a similar scenario but this time with 7 expert fake news annotators. We present the description of both studies, characteristics of the resulting dataset (which we hereby publish) and several findings.
\end{abstract}


\section{Introduction}

% Fake news, as species of disinformation~\citep{gelfert2018fake}, are understood as news articles that are intentionally and veritably false and could mislead readers~\citep{allcott2017social}. 

Fake news is a form of misinformation~\citep{gelfert2018fake}. They can be understood as news articles that are intentionally false and could mislead readers~\citep{allcott2017social}. 
Fake news (and misinformation in general) have gained notoriety in recent years for spreading in online environment, particularly via social networks~\citep{Fernandez2018, Kumar2018}. They have impact on real-world human decision making, for example in politics~\citep{Balmas2014, Garrett2011}. They are also behind cases of \emph{offline harm}, such as in the health-care domain~\citep{Kata2010,gualtieri2009}.

The real and significant negative impact of fake news spreading has prompted intensive research efforts. Some studies have been directed towards understanding of the fake news phenomenon (i.e., \emph{characterization}). They analyze the properties of the fake news~\citep{Kumar2018, Shu2017, Parikh2018}, how they gain trust of the readers~\citep{Heuer2018}, how fake news spread and what drives their spreading~\citep{Vosoughi2018, Flintham2018}. The existing studies have helped greatly in conceptualizing the fake news. However, \emph{these studies call for yet more behavioral studies to investigate the factors of human judgment during the interaction with (fake) news~\citep{Vosoughi2018}}.

Building on the knowledge from characterization studies, automated fake news detection approaches~\citep{Shu2017, Parikh2018, Conroy2015a} and platforms~\citep{Saez-Trumper2014} have been proposed. The prevailing trend is to use machine learned approaches. Some methods analyze the news textual content~\citep{Ajao2018, Parikh2018}, some focus on graphics~\citep{Jin2017a, Gupta2013}. Others rely on contextual aspects, such as source/author/spreader models~\citep{Boididou2018} or social media interactions~\citep{Volkova2017}. Despite such efforts, the reliable \emph{detection of online misinformation remains an unsolved issue}. There is a general lack of datasets that would cover the problem in all its heterogeneity. Also, the dimensionality of news content and context is high. This makes the creation of general and reliable classification models difficult.

%The veracity estimation from news content involves extraction of true meaning of texts, where even little details can make the difference. The use of contextual features is very helpful, but not reliable on its own.

To address some of the issues mentioned above, we present two user studies that investigate the human behavior during fake news consumption and veracity evaluation. In the first, main study, 44 lay participants first casually consumed news stories in a social-network-like feed (i.e., the \emph{consumption phase}). After that, they were asked to estimate the veracity of these stories (i.e., the \emph{evaluation phase}). In the second, follow-up study, we replicated the evaluation phase of the first study, but this time with 7 experienced fact-checkers (dubbed the ``experts''). The motivation for such studies has two core reasons:
\begin{enumerate}
    \item \emph{Understanding the fake news consumption.} There is a general lack of understanding of how people consume news on social media with respect to the fake news phenomenon. By observing the news consumption behavior, we may disclose what exactly draws the attention of news readers in case of fake news, how thorough they are in consuming the news, what precedes their explicit actions (e.g., likes or shares) or under what conditions they become ``spontaneously'' suspicious about a news piece.
    \item \emph{Learning from human fact-checkers.} High dimensionality of the misinformation detection task can possibly be eased by smart selection of relevant features. Observing the study participants' behavior during evaluation phase can point us to content and context features of online news, which are important for veracity validation. This particular motivation prompted us to execute the follow-up study, as we wanted to cover the behavior of expert annotators as well (the main study comprised only lay participants with no known experience with fake news annotation).
\end{enumerate}

With this work, we aim to fill in a blank spot in the area of fake news consumption and detection. Additionally, our study is complemented with the use of eye-tracking equipment for a more detailed recording of participant behavior.

% (too strong forumulation) to the best of our knowledge, \emph{no eye-tracking study of fake news consumption and veracity evaluation has ever been conducted}.

This work has the following contributions:
\begin{enumerate}
    \item \textit{The concept of the fake news reading study.} We present the study concept as a separate contribution to the state-of-the-art. Firstly, we are not aware of any previous study done in such setting. Secondly, we recognize the increasing demand for detailed recordings of social network user behavior and thus, we wish to provide a blueprint for future studies.
    
    \item \textit{The dataset of 44 recorded participants of our study.} With this paper, we make the dataset gathered in this work publicly available\footnote{http://bit.ly/fake-news-reading-dataset}. In section~\ref{sec:findings}, we provide a descriptive analysis of the recorded behavior data. We also report on the results of a questionnaire, which we issued among the participants, which gathered reference information on participants' opinions and degrees of interest in topics featured in the news.
    
    \item \emph{The findings on consumption and veracity evaluation of social media users.} Using the data obtained during the consumption phase of the experiment, we investigate and report the influences that the fake news have on the user behavior within the social media website feed. By analyzing the data recorded during the veracity evaluation, we report on strategies employed by successful annotators.
\end{enumerate}

\section{Related Work}

Several studies investigating human behavior during fake news consumption or verification have been conducted in the past. We can draw from them methodologically and complement their findings by our study.

%To the best of our knowledge, there are no \emph{eye-tracking} studies that examine human behavior during fake news consumption or verification. There are, nevertheless, several similar \emph{non-eye-tracking} studies. We can draw from them methodologically and complement their findings by our study.

From the existing research, the most similar to our study is the recent work of~\cite{Flintham2018}, who investigated the fake news reading behavior of social media users. First, they surveyed 309 questionnaire respondents for experience with fake news and their characteristics. Second, they conducted a qualitative study with 9 participants, who interacted with a fake Facebook account feed, which contained true and fake stories of various kinds using the typology of~\cite{Rubin2015}. The authors investigated, to what extent the social media users are concerned about the spread of fake news and what (content or context) features convince them to believe that an information is fake. Using qualitative analyses based on think-aloud protocol, Flintham et al. identified three general strategies used by the news readers during truthfulness assessment. The first approach was to infer the truthfulness solely based on the source reputation. Second was to assess the story content and then confirm the assumption with the source reputation. Third strategy relied on the content only. When analyzing content, participants mostly relied on plausibility assessments and headline and writing style analysis. 

Our own studies, presented in this paper, are complementary to the work of~\cite{Flintham2018}. Our aim was to involve a larger set of participants (44 compared to 9) and analyze their behavior in a quantitative manner. Our study protocol was silent (as opposite to think-aloud), as we wanted to emulate more authentic situations. We have also employed eye-tracking to track the reading and decision-making behavior in detail. The environment in our study is more controlled: rather than creating a fake account on real Facebook, we created an entire application imitating the Facebook feed. Furthermore, as the story source (and its reputation) is evidently a strong indicator of truthfulness for some users, we have purposefully omitted the sources of articles used as stories. Also, when participants in our study decided to open the full articles, the content was presented in a unified graphical way. Therefore the original source could not be inferred that way either.

In a different study, Heuer and Breiter examined factors that influence the \emph{trust} of users towards news shared on social media~\citep{Heuer2018}. They worked with young adults, who represent one of the groups mostly threatened by misinformation. This group primarily accesses news online and on social media: 53\% of users aged 18--24 reported to access news on social media, according to a recent Reuters study~\citep{reuters2018}. In our main study, we chose the same demographic group. However, in contrast to Heuer and Breiter, we do not evaluate the trustworthiness of the news source, nor other social media signals, such as number of likes or comments (which Heuer and Breiter did not find significantly influencing the users' assessments).

\cite{Torres2018} proposed a model of news verification behavior, which they verified by the means of a survey administered to 541 participants (university students; 64.2\% were aged 19--24). They found media credibility, users' intention to share, their fake news awareness, and their trust in the network as relevant predictors of news verification behavior. However, these findings were not further evaluated in an experiment or a user study.

Several existing eye-tracking studies were focused on news reading behavior in general. \cite{Arapakis2014} examined what interests the readers of online news. The authors conducted an eye-tracking study, in which they measured engagement in 114 articles presented to 57 participants. The articles had different properties regarding their sentiment and polarity, which indeed influenced their attractiveness and consequently, the reader behavior. The work of Arapakis et al. also has some implications for our studies. The behavior should be investigated over multiple topics found in the news, as the personal long-term interests of participants can influence their consumption as well as veracity evaluation behavior. The topical interests and opinions of participants should be controlled through questionnaires.

Lastly, there have already been some works on identifying indicators of credible vs. not credible news. Our study presented in this paper also aims to tackle these by analyzing strategies of successful users, i.e., users who can detect fake news with high accuracy. \cite{Zhang2018} proposed a set of content and context indicators (features) which they evaluated on a dataset of 40 articles with human annotators and domain experts. From the content features, clickbait title and a presence of a logical fallacy (e.g., a slippery slope) turned out to be good predictors of the credibility of an article. From context factors, the good predictors were the outcome of the fact-checking (article being flagged as false), number of social and mailing list calls (i.e., calls to share the article or to subscribe to a mailing list) and placement of ads and social calls (i.e., how aggressive these were). However, the automatic assessment of some of these features can be a problem and the authors do not discuss how to overcome it.
\section{Study description (main study)}
\label{study}

To gather detailed data on how social media users perceive and consume fake news, we have conducted a live, controlled study with 44 participants. We first asked the participants to browse (and optionally read) news-like stories (posts) in an experimental user interface resembling a social media feed. This, we denote as the consumption phase or \emph{the first pass}. Then, we asked the participants to browse the same stories again. This, we denote as the evaluation phase or \emph{the second pass}. This time, the participants had to determine, according to their own judgment, the veracity (truthfulness) of the stories. Each pass took exactly 12 minutes and featured 50 stories, some of which were true and some fabricated. Each story belonged to one of the 11, rather controversial, topics.

Throughout the study, the behavior of participants in the social media feed was recorded in detail, including the use of eye-tracking. In addition to this, we have issued a questionnaire that surveyed for degrees of interest and opinion positions of participants in each of the 11 topics.

\subsection{Research questions}

The study was motivated by the following general research questions:
\begin{description}
    \item[RQ1] What behavior traits characterize the consumption of fake news in a social network feed environment? 
    \item[RQ2] How do the long term interests and opinions of participants influence the consumption behavior?
    \item[RQ3] What behavior traits characterize the participants that were relatively successful in veracity evaluations? 
    \item[RQ4] Are there some apparent content features, which successful annotators use for their evaluations? Are they potentially usable in automated detection approaches?
\end{description}
The study has largely an exploratory character. We did not set up any hypotheses upfront.

\subsection{Participants}

The experiment was attended by 44 participants (27 males and 17 females). The mean age of the participants was 17.5 (SD 0.8, min 16, max 20)\footnote{Most of the students were in the last year of their high-school studies, which, in Slovakia, roughly corresponds with the age of 18.}.

All of the participants were high school students. We have selected this particular demographic group for two reasons. The first reason was the effort to reduce the potential confounds. It would, of course, make sense to try to cover the entire demographic spectrum by inviting a representative sample. This would, however, introduce many uncontrolled variables and potential confounds into the study (such as variability in information literacy in different demographic groups). Thus, we narrowed the study to a specific, more homogeneous group -- the high school students.

The second reason for narrowing down to a specific group was that we wanted to work with a group somewhat susceptible to misinformation online, similarly as~\cite{Heuer2018}. In Slovakia, where the study has taken place, the high school students represent such a group~\citep{STRATPOL2018}. There were also other specific groups worth inviting. For example, elderly people are also known for their susceptibility to misinformation. The choice of high school students was, in this case, motivated by a better practical availability of participants.

Despite the use of a specific demographic group, we deem the study design usable for other groups or general population sample as well.

\subsection{Environment and equipment}

The study recording was executed in the laboratories of the User Experience and Interaction Research Center\footnote{http://uxi.sk} at Slovak University of Technology in Bratislava.

The recording was done in a group eye-tracking laboratory, which means that multiple participants attended the study and were recorded in parallel. This method of eye-tracking was closely described by~\cite{Bielikova2018}. While this setup is not required by the study design, we used it to execute the study faster. The group eye-tracking studies require specific scenario adjustments for group moderation and are slightly more demanding for servicing personnel. They also require special infrastructure to support recording and collection of data. The investment is, however, worthwhile: this study's recording part was done in about 4 hours: in 4 one-hour sessions, including overheads. If the same study was to be done with a sequential setup, it would take more than 40 hours.

The participants were working on desktop computers with 24 inch monitors (1900x1200), mounted with Tobii X2-60 eye-trackers. The experiment was run using Tobii Studio software (for the first and second phase). The questionnaire (third phase) was issued through Google Forms. The workstations were placed in a room with stable lightning conditions. The layout of the room is classroom-style with 20 workstations. Maximum of 13 workstations was occupied during a single session. The environment was silent during the 12 minute passes: instructions were given orally before, after and in between the passes.

\begin{figure}[t]
    \centering
    \includegraphics[width=\columnwidth]{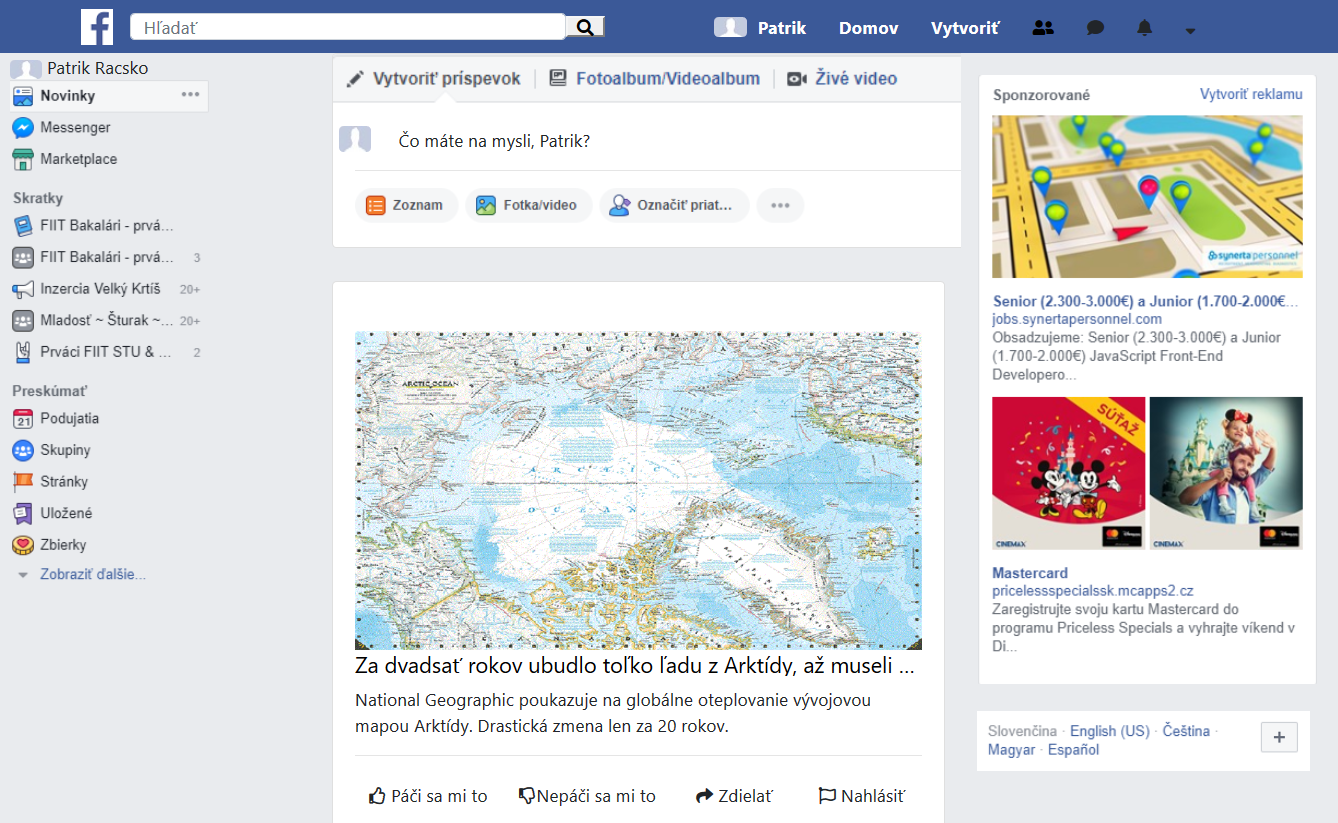}
    \caption{For the purpose of the experiment, we have created a website resembling the real social network user interface.}
    \label{fig:interface-example}
\end{figure}

\begin{figure}[t]
    \centering
    \includegraphics[width=\columnwidth]{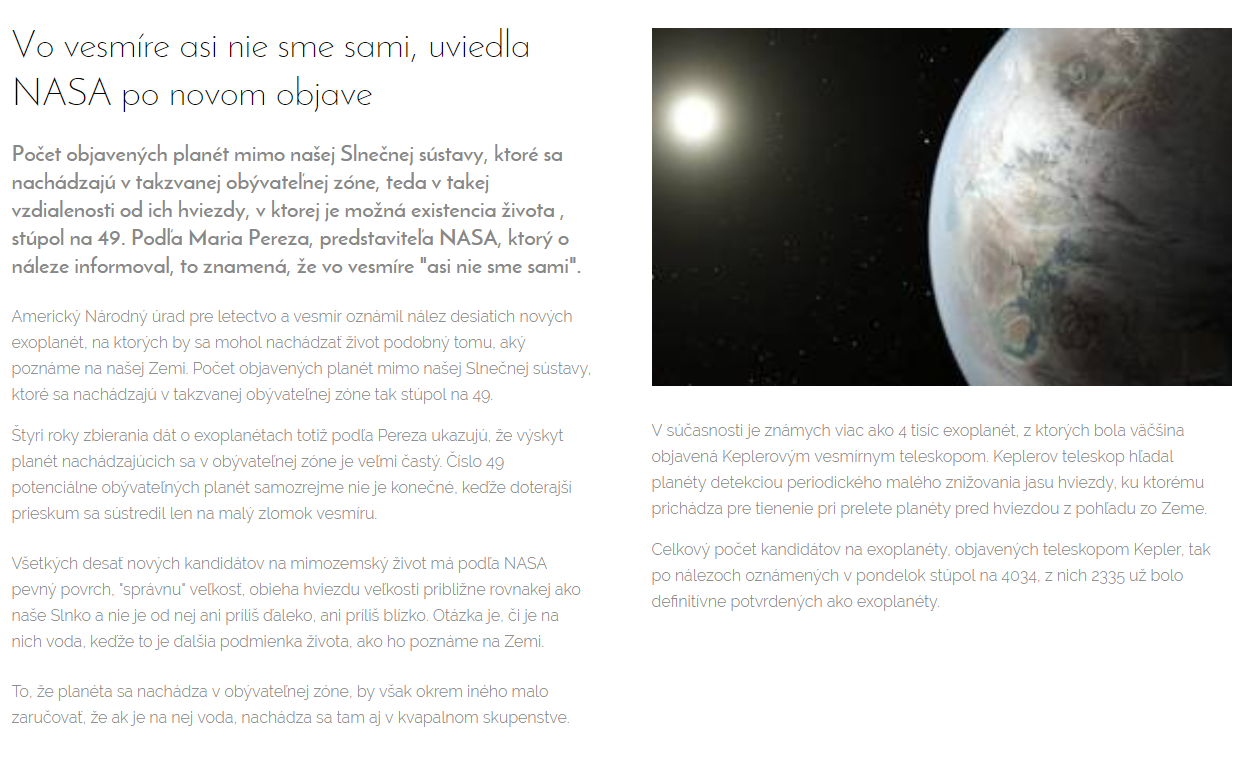}
    \caption{An example of a full article (displayed when a participant decides to open it from the feed). All articles were set up in the same layout and style.}
    \label{fig:article-example}
\end{figure}

For the purpose of the study, we have created a website that imitates the real social network (Facebook) user interface. The website comprises a ``post feed'' (see figure~\ref{fig:interface-example}), through which the participant can scroll and which also allows limited interaction options (e.g. liking, sharing in the first phase and veracity evaluation in the second phase). A participant may also open up the full article (see figure~\ref{fig:article-example}) by clicking on the picture, title or teaser text in the feed post.

\subsection{Study scenario (tasks)}

Each session started with the arrival of the participants to the group eye-tracking laboratory. First, we introduced them to the study: \textit{``This study investigates the reading habits of people on social network sites ...''}. Next, they were briefed about the workings of the eye-tracking technology and the eye-tracker calibration was performed with each participant. 

\begin{figure}[t]
    \centering
    \includegraphics[width=\columnwidth]{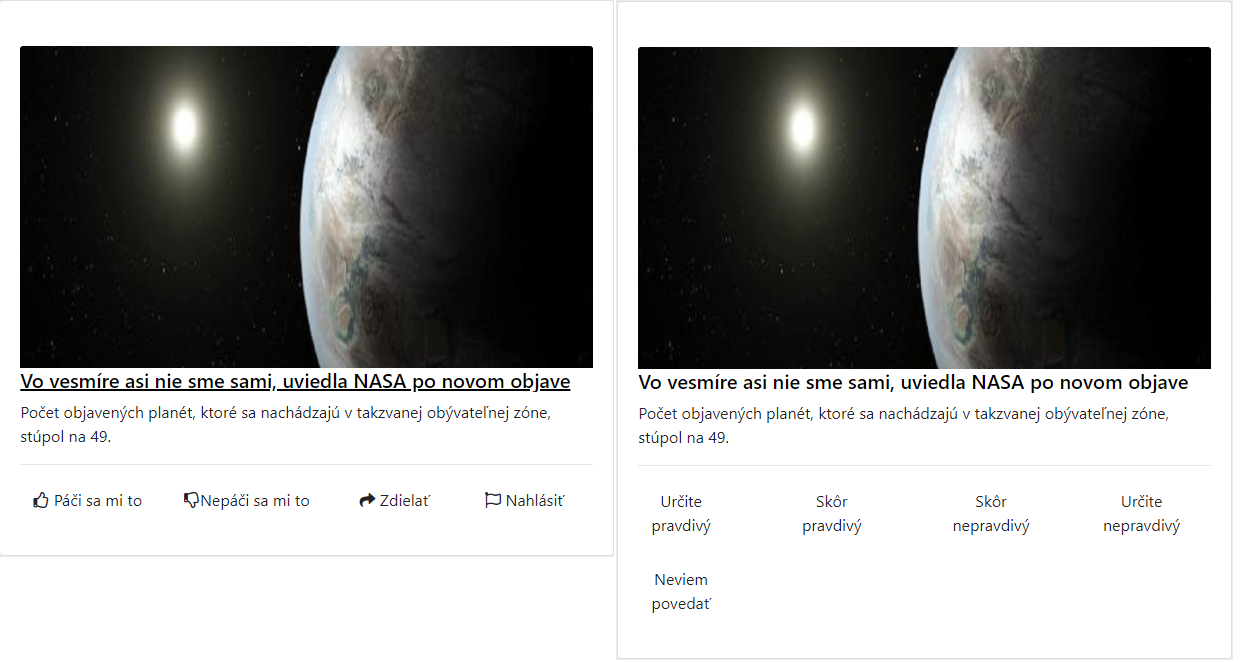}
    \caption{\emph{Left:} example of a social network post during the first task (news consumption phase). A participant could like, dislike, share or report the article. \emph{Right:} the same post during the second task (veracity evaluation phase). A participant could indicate his/her estimation of veracity (options: definitely true, probably true, probably false, definitely false and the option: I don't know in the second line).}
    \label{fig:post-examples}
\end{figure}

Then, the participants were asked to perform the following tasks:

\begin{description}
    \item[Task 1:] \textbf{News consumption (the first pass).} The participants were asked to casually browse/read the social network feed and the articles to which the posts were linked. The task lasted 12 minutes and its purpose was to record the news consumption behavior in a situation resembling the casual ``checking out'' of the social network feed. The key part of the instructions was formulated as follows: 
    \emph{``We will display an imitation of a social network interface to you. For 12 minutes, casually browse the story feed. You may open up the full articles. You may also like, dislike, share or report any story -- do so if you feel you would do it in a real situation.''} Furthermore, we restricted the activity of participants to our experimental application (we asked them not to visit other websites).
    
    \item[Task 2:] \textbf{News veracity evaluation (the second pass).} We informed the participants that some of the articles in the feed are fake news. We asked them to look at the feed again and estimate the veracity of articles based on their content (either in the feed post or in the full version). Note, that up to this point, the participants were not hinted in any way, that the true purpose of the study is to investigate the fake news reading behavior (we were careful not to spoil their behavior during the task 1).
    The key part of the instructions was formulated as follows: 
    \emph{``You will again see the same articles in a similar environment. This time, however, your task will be different. Your goal is to determine, whether a given article is true or not. Some of the articles in the feed are not based on truth.''}
    We anticipated that the veracity evaluation can take some time for each article. Therefore we further instructed the participants: \emph{``Please, evaluate the articles at your own pace, it is not necessary that you evaluate all of them.''}
    The task lasted for 12 minutes.
    The participants indicated the veracity using one of the five options available at each post: \emph{certainly true, rather true, rather false, certainly false, I don't know} (see figure~\ref{fig:post-examples}). 
    
    \item[Task 3 - questionnaire] We asked the participants to fill out the questionnaire for assessing demographics, opinion positions and interests for featured topics. The questionnaire is described in section~\ref{sec:questionnaire}. This task was not eye-tracked.
\end{description}

All of the instructions were given orally to the entire group at the same time. The instructions regarding tasks were always given immediately before a task started. A task was always explained and we gave the participants a chance to ask questions before the task started. This ``synchronized'' style of study organization was implied by the group eye-tracking setup. The first two tasks had a fixed time length (12 + 12 minutes). During the third task, we let the participants to work at their own pace. The entire session (there were 4 of them) took between 35 to 40 minutes.

\subsection{Stimuli (content characteristics)}

In both first and second task, the feed displayed to the participants contained 50 posts. Each post represented one article. The full articles had length of approximately 350 words, so the entire text would fit the screen when the full article was opened up (see figure~\ref{fig:article-example}). Each full article had a title, an image that illustrated it, a lead paragraph and the rest of the text. The post representation of the article (in the feed) featured the same title and image and the beginning of the lead paragraph (see figure~\ref{fig:post-examples}). The lead paragraph was truncated, if it did not fit the two lines allocated in the post interface. Note, that we have purposefully omitted the sources and authors of the articles and also opted for unified visual setup in the full article view (i.e., all articles looked visually the same).

The articles were written in Slovak language (all of the participants were native Slovak speakers). The style of writing resembled news or magazine articles and were expected to be easily understandable by the participants.

Each of the articles belongs to one of the 11 \emph{topics} (content categories), hand-picked for this study: \textit{food quality, personal weight reduction, Muslims in the EU, pharmaceutical industry, LGBT rights, right wing extremism\footnote{The articles in this topic were centered around a particular political figure in Slovakia.}, vaccination, extraterrestrial life, global warming, deforestation, migration}.

Within the societal context of Slovakia, the presence of fake news (or misinformation, in general) is more or less common for each of the topics we used.

Each of the 11 topics also bears some degree of controversy. In each of the topics, a dominant \emph{dimension of dispute} can be found, i.e., there are two general opposing opinion positions. For example, in the topic of \emph{global warming}, we can recognize the opposing positions of a \emph{denier} and an \emph{alarmist}.

For each category, we gathered 4 articles to fill each of the combination of veracity and opinion position. This means that we always had at least one true article supporting each opinion position and at least one fake article supporting each opinion position.

We used the following definition of a fake article: it must contain at least one false \textit{factual claim} or it must support a \textit{narrative} commonly known as false. However, such definition is quite extensive and could contain many edge-cases. Thus, when we prepared our experimental data, we preferred articles with multiple false claims and false narrative support. The fake articles we used were also not fully false, each contained some truthful claims as well.

As examples, we can list the titles of 4 articles from the \emph{global warming} topic:
\begin{itemize}
    \item \textbf{true-alarmist:} \textit{``So much ice has disappeared from the Arctic, that the world atlas needed to be updated''}
    \item \textbf{true-denier:} \textit{``Japanese scientists: the global warming is not caused by humans''\footnote{This particular article was true, although it referenced to a rather old study.}}
    \item \textbf{fake-alarmist:} \textit{``A published study describes the catastrophic consequences of warming''}
    \item \textbf{fake-denier:} \textit{``DiCaprio had sold himself to a poor propaganda, saying that the climate change is the biggest threat to the future of our society''}
\end{itemize}

We have gathered most of the articles directly from well-known Slovak news websites and (in case of fake news) from notorious Slovak misinformation sources. The original articles had varying length (usually more than we needed). We shortened some of them by omitting paragraphs exceeding the required length. For some combinations of topics and opinion positions, it was impossible to find examples of fake news. For example, we found none fake news articles that would \emph{support} the vaccination cause. In such cases, we created the required fake news articles ourselves, usually by changing existing articles using exaggeration (e.g., ballooning the number of measles victims 10 times), fact changing (e.g., changing the names of notoriously known protagonists) and image and title changing (so the articles were more suggestive). 

In total, we used 50 articles in the experiment. This means that for some combinations of topic, veracity and opinion position, there was more than one example. In total, this happened in 6 cases. The reason for including more articles was, that we sometimes came across multiple interesting examples (usually fake news) that were distinct to each other in some sense and we wanted to record behavior of participants over both of them.

Each participant of the experiment was presented with all 50 articles. The order was randomized for each participant, but that order was then kept the same between first and second phase of the experiment. The randomization was done to remove potential confounds caused by unwanted priming effects.

\subsection{Questionnaire}
\label{sec:questionnaire}

The main purpose of the questionnaire was to acquire the information on \emph{opinion positions} and \emph{degree of interest} of participants for each of the 11 topics. We have issued this questionnaire as the third (and last) task of the study to avoid priming effects on the participants' behavior in tasks 1 and 2.

The questionnaire had the following structure:
\begin{enumerate}
    \item Questions on demographics (gender and age).
    \item Questions on \emph{opinion positions}. These were organized in 11 sets (one for each topic). Each set contained between 5 to 7 specific questions.
    \item Questions on \emph{degree of interest} (one question for each topic).
\end{enumerate}

For acquiring the opinion positions of participants within the dimensions of dispute, we opted for multiple specific questions (rather than asking one general question to acquire the position as a whole). We did so, because we were afraid that the participants would not able to directly evaluate themselves.

The questions on opinion positions had the form of statements to which the participants expressed their agreement/disagreement using a five-level Likert scale: \emph{(+2) definitely agree, (+1) somewhat agree, (0) cannot judge, (-1) somewhat disagree, (-2) definitely disagree}.

As an example, we can take the set of statements corresponding to the \textit{global warming} topic:
\begin{enumerate}
    \item \emph{``The global warming is a fabrication and is not really happening.''}
    \item \emph{``The global warming is a serious threat for our entire planet.''}
    \item \emph{``People cause the Earth warming.''}
    \item \emph{``Personally, I do not contribute to the global warming.''}
    \item \emph{``The global warming has not, and will not have any effect on me.''}
\end{enumerate}

As can be seen, a statement has always its own opinion polarity in itself and can be somewhat suggestive. Therefore, in each set of questions, we included statements directed both ways. In the example above, the statements no. 1, 4 and 5 were directed towards the \emph{denier} position, while statements no. 2 and 3 to the \emph{alarmist} position. The overall opinion position was then computed as average answer for all statements in the set. Before that, however, the contributions of statements biased towards one of the positions (arbitrarily chosen) were multiplied by -1. In the example above, we could either ``flip'' the contributions of statements 2 and 3 or statements 1, 4 and 5.

Regarding the questions on degree of interest, these were formulated one per each topic as follows: \emph{``Select your degree of interest in the following topics''}. This instruction was followed by the list of 11 topics, each assigned with a 5 degree scale ranging from `\emph{`I'm not interested at all''} to \emph{``I'm very interested''}.

\subsection{Data processing}

After recording, we used Tobii Studio to compute fixation gaze events using the built-in implementation of IVT fixation filter. We also used Tobii Studio to define areas of interest (AOIs) over each post in the feed: image, title, perex (leading paragraph text), action buttons (like, share, etc.) and one AOI denoted the entire post (including all previous AOIs). The fixation data (with AOI hits) were exported and further processed using Python scripts.

\section{Findings (main study)}
\label{sec:findings}

% - deskriptivna analyza ✓
%    - cas na feede/ v clankoch v K a E ✓
%    - pocet precitanych ✓
%    - pocet explicitnych akcii v K ✓
%    - pocet ohodnotenych clankov ✓
%    - pocet videnych postov v K a E ✓
%    - analyza z dotaznika per tema ✓
%        - nazory, zaujem a polarita ✓
% - findings v konzumacii obsahu
%    - explicitne akcie
%        - su konzistentne s hodnotenim pravdivosti?
%        - suvisia s hodnotenim pravdivosti?
%            - typ reportu vs pravdivost pocty
%    - suvis zaujmu a dlzky citania clankov v K ✓
%        - dlzka citania full (otvorenych) clankov ✓
%        - pokrytie citania full clankov ✓
%        - cas straveny v poste na feede ✓
%    - pomery obrazky v nadpisy v perex vo feede (deskriptivne) ✓ -- rozdiely su medzi prvou a druhou castou
%    - atraktivnost nadpisov, obrazkov v pravdivych vs fake clankoch %vo feede
%        - paired t-test ✓
%        - voci deklarovanemu zaujmu ✓
% - findings o spravani hodnotitelov
%    - uspesnost per participanta
%        - korelacia s ich nazormi
%        - pozriet sa na najuspesnejsich
%            - co si vsimali v com boli odlisni
%            - rozdiely v spravani
%            - eye tracking metriky
%            - pocet citanych full clankov ✓

%TODO sem davat findings

%This section summarizes the main findings from the study. It first gives key descriptive statistics

\begin{figure*}[t]%
    \centering
    \includegraphics[width=\linewidth]{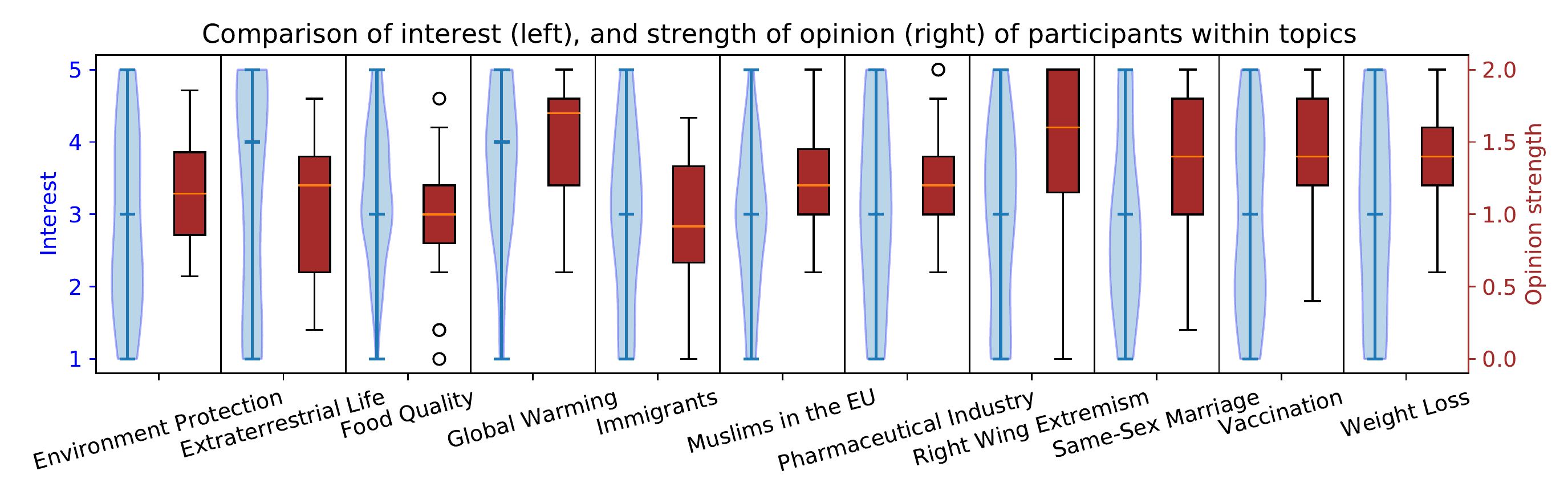}%
    \caption{
        Comparison of distributions of strength of interest and strength of opinion of participants within topics based on responses in final survey.
        Strength of opinion is calculated based on agreement with statements in the survey.
        Strong agreement or disagreement is represented using the value 2 and lack of opinion as 0.
    }%
    \label{fig:topics-opinion-interest}%
\end{figure*}

\subsection{First and second pass statistics}

Participants spent on average 5.1 (SD 2) out of 12 minutes browsing the feed in the first pass.
On the second pass, they spent 5.3 minutes (SD 2.6) on average out of 12 minutes browsing the feed.
The remaining time was spent reading full articles.
The participants opened on average 7 articles (SD 4) in the first pass and 9 (SD 5) in the second pass.

In the first pass of the experiment, participants saw on average 41 posts in the feed (SD 10). We define the ``seeing of a post'' (impression) as a situation in which we record at least 10 fixation events hitting the area of interest denoting the post. This normally means couple of seconds and by imposing this, we wished to filter out fast scrolling through the feed and other possible artifacts.

22 of the participants performed a total of 122 explicit like, dislike, report, and share actions.
An equal number of actions (61) were performed both for truthful as well as fake articles, however the distribution of the actions for truthful and fake articles differed.
Resembling the intended use of the actions, truthful articles received more likes and shares than fake articles, and fake articles were disliked and reported more often than truthful articles.
Table~\ref{table:actions-first-pass} shows the numbers of explicit actions in the feed separately for truthful and fake articles. The total number of explicit actions was low and their distribution skewed heavily towards few participants (majority of explicit actions was done by just 4 participants). Therefore, we cannot draw many conclusions from the explicit behavior records. However, we can see some potential in the report action that was used far more for the fake articles than for truthful ones.

\begin{table}[ht]
\begin{center}
\caption{Number of actions recorded for truthful and fake articles in the first pass.}
\label{table:actions-first-pass}%
\begin{tabularx}{0.85\columnwidth}{lXXXX}
    \toprule
    &  Dislike &  Like &  Report &  Share \\
    \midrule
    True Articles       &       18 &    43 &       2 &      7 \\
    Fake Articles       &       21 &    34 &      10 &      5 \\
    \bottomrule
\end{tabularx}
\end{center}
\end{table}

On the second pass of the experiment, the participants were somewhat slower in passing through the feed: they saw on average 32 posts in the feed (SD 13).
Of those, they rated the veracity of 29 articles on average (SD 16).
Table~\ref{table:actions-second-pass} shows the total numbers of veracity ratings for truthful and fake articles.
There is a statistically significant difference in the distribution of veracity ratings for fake and real articles (unpaired t-test of ratings per participant and topic results in p-value $<$ 0.001).
Given a scale of 1 to 5, where 1 stood for certainly true veracity rating, the average veracity rating of truthful articles was 2.5 (SD 1).
The average veracity rating of fake articles was 3.4 (SD 1).

\begin{table}[ht]
\begin{center}
\caption{Number of ratings made for truthful and fake articles in the second pass.}
\label{table:actions-second-pass}%
\begin{tabularx}{\columnwidth}{lXXXXX}
    \toprule
%    &  \thead{Certainly \\ True} &  \thead{Rather \\ True} &  \thead{Can't \\ Tell} &  \thead{Rather \\ Fake} &  \thead{Certainly \\ Fake} \\
   &  Certainly True &  Rather True &  Can't Tell &  Rather Fake &  Certainly Fake \\
    \midrule
    True Articles       &               134 &             214 &                 77 &           131 &              45 \\
    Fake Articles       &                65 &             172 &                 72 &           174 &             206 \\
    \bottomrule
\end{tabularx}
\end{center}
\end{table}

\subsection{Opinion and interest in topics}

Responses from the survey enabled us to explore distributions of interest, strength and polarity (position) of opinions of participants within the relevant topics.

Figure~\ref{fig:topics-opinion-interest} compares the distribution of strength of interest and strength of opinion of participants. The \emph{interest} represents the interest in a topic as explicitly stated in the last part of questionnaire. On the other hand, the \textit{strength of opinion} represents the extremity of answers to the opinion position questions.

\begin{figure}[ht]%
    \centering
    \includegraphics[width=.7\linewidth]{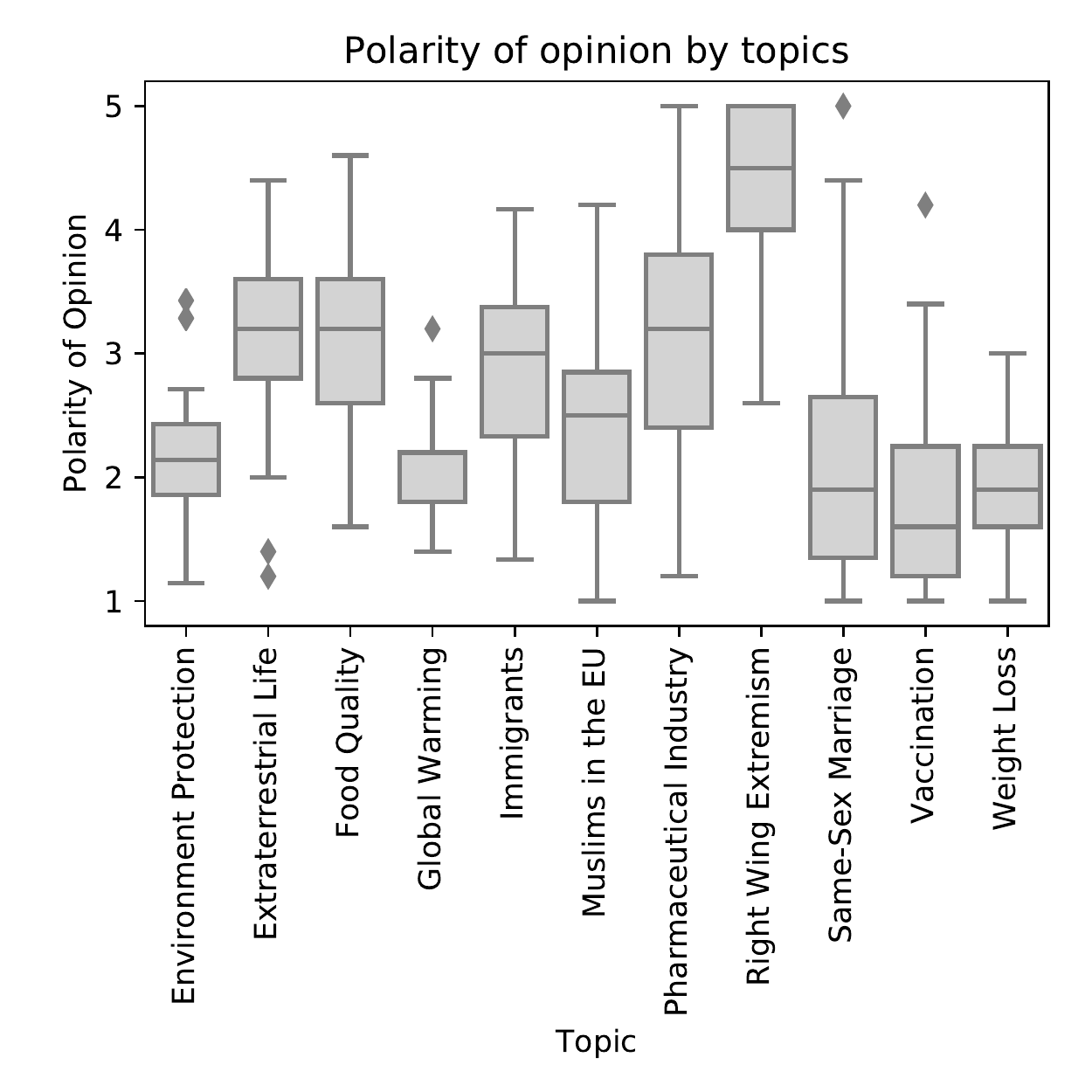}%
    \caption{
    Polarity of opinion calculated based on agreement with statements in the survey. Agreement with statements with opposing polarity is represented as disagreement.
    In the shown distributions, there are both topics where the participants generally leaned towards a single end of the polarity spectrum (e.g., right wing extremism) as well as topics where both polarities were represented (e.g., pharmaceutical industry).
    }%
    \label{fig:opinion-polarity-topics}%
\end{figure}

In most topics, strength of opinion matches interest of participants which is an expected outcome.
Yet, there are several topics where the distribution of strength of opinion is relatively higher compared to the distribution of strength of interest---e.g., \emph{right wing extremism}, \emph{same-sex marriage}, and \emph{vaccination}.
In these topics, we see that the participants lean towards one end of the polarity spectrum---they are either leaning towards strong agreement or strong disagreement with issues presented in the topic.
Figure~\ref{fig:opinion-polarity-topics} shows the distribution of opinion polarity ratings within topics, where the described topics seem to be more skewed toward one extreme compared to more uniform distributions of other topics.
We cannot reliably interpret this observation.
One suggestion may be that the opinions of participants were shaped by their external environments.
This is suggested by the difference between distributions of interest and strength of opinion ratings and the agreement within ratings of polarity.

%Such topics in which the participants exhibit strong opinions with relatively consistent polarity while lacking interest suggest that the opinions of participants were shaped by their external environments.
% TODO prehodnotit ci tento argument chceme nechat

We further studied the relation between explicitly stated interest of participants and their implicit interest observed from their behavior during article reading in the first pass (\emph{consumption phase}) of the experiment.
Compared to participants with stated disinterest (rather uninterested, not interested at all),
participants with stated interest (rather interested and very interested)
opened more articles (average 0.8, SD 0.9 compared to 0.3, SD 0.7), and
spent more time reading them (average 1.5 minutes, SD 1 compared to 1.3 minutes, SD 1.2).
This finding shows that the implicit reading behavior of participants in the first pass of the experiment reflected their explicitly stated interest, i.e., participants followed their interests when exploring the feed in the consumption phase of the experiment.

\begin{figure}[t]%
    \centering
    \includegraphics[width=.7\linewidth]{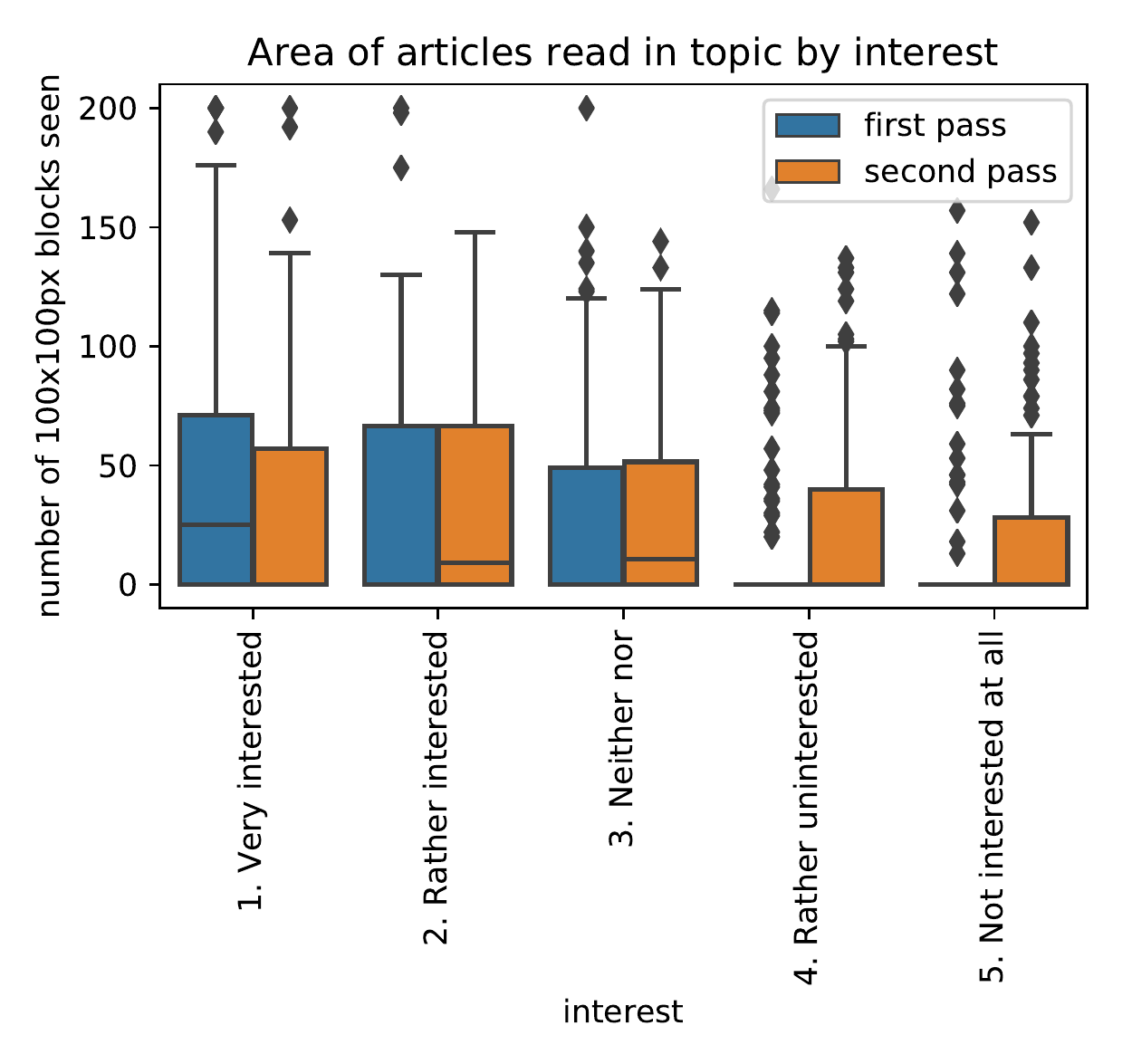}%
    \caption{Total area of articles within topics seen in the first and the second pass of the experiment in relation to the declared interest of participants. The influence of interest on the total seen area of articles is much higher in the first pass compared to the second pass.}%
    \label{fig:interest-read-area}%
\end{figure}

On the other hand, the reading behavior in the second pass of the experiment was less influenced by the interest of participants.
Compared to participants with stated disinterest, participants with stated interest
opened only slightly more articles (average 0.8, SD 1 compared to 0.7, SD 1) in the topics.
Figure~\ref{fig:interest-read-area} shows the difference in the influence of interest in topics on the total seen area of articles in the first and second pass.
The total seen area of articles reflects the depth in which participants read opened full articles (as recorded in gaze data). 
In contrast with the first pass of the experiment, the participants read articles in topics that they were not interested in more deeply when tasked to rate their veracity.
Nevertheless, an influence of interest on the reading behavior in the second pass can still be observed in the distribution of the total seen area in topics. This somewhat differs with the findings of Flintham et al., who, on the contrary, observed, that stories in which the participants were not interested ``mattered little to them despite repeated prompting'' to evaluate their veracity~\cite{Flintham2018}.

\subsection{Correctness of veracity ratings}

\begin{figure*}[htb]%
    \centering
    \includegraphics[width=\linewidth]{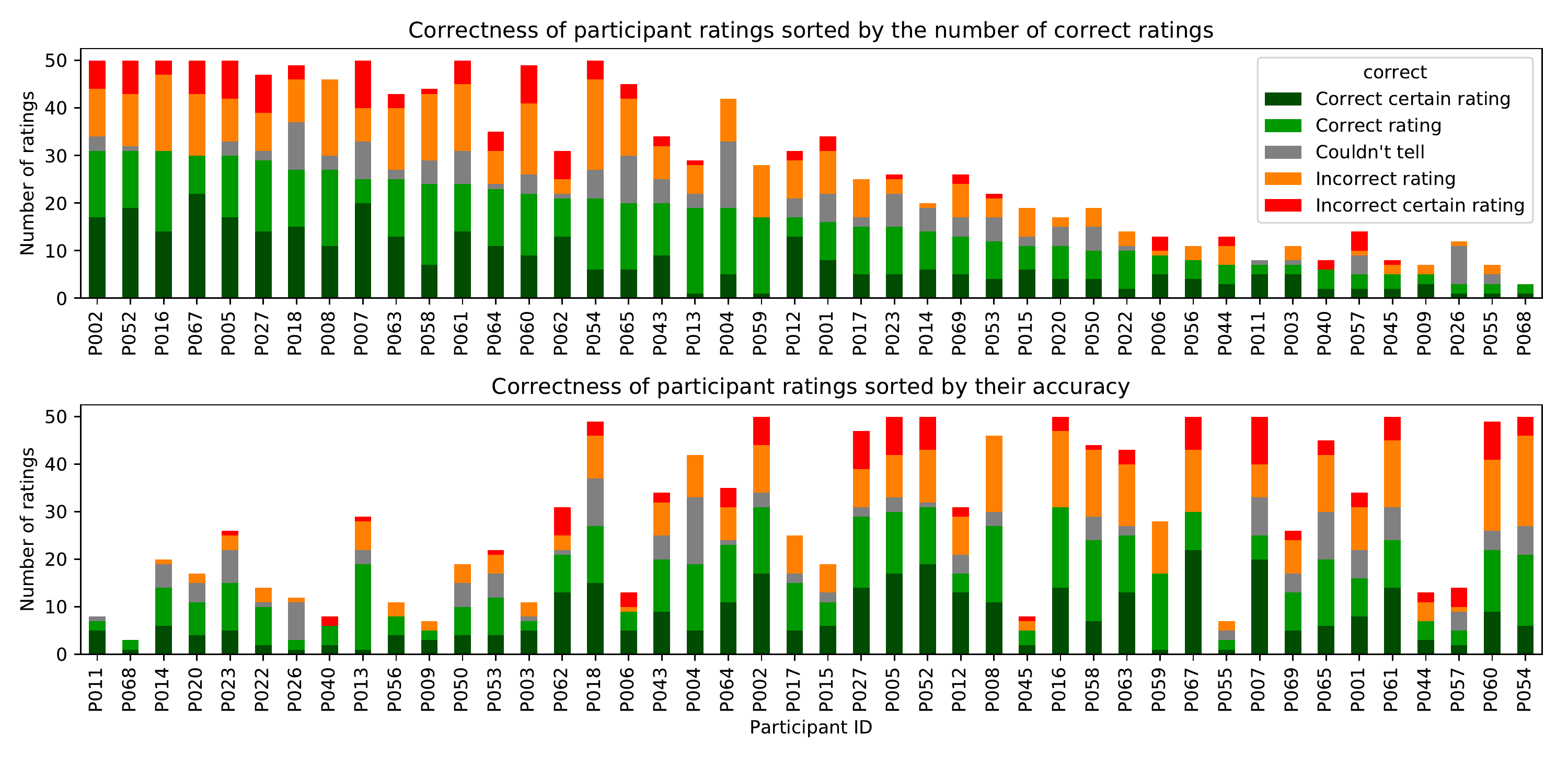}%
    \caption{Distribution of veracity ratings for each participant. In the top chart, participants are sorted by the number of correct ratings. The bottom chart shows participants sorted by the accuracy of their veracity ratings. Sorting by accuracy gives an almost mirror order of participants than sorting by number of correct ratings, which suggests that participants who spent more time rating individual articles achieved higher accuracy.}%
    \label{fig:ratings-by-participants}%
\end{figure*}

Figure~\ref{fig:ratings-by-participants} shows distributions of veracity ratings per participant sorted by the number of correct ratings (top) and the overall accuracy of each participant's ratings (bottom).
Sorting the participants by accuracy gives an almost mirror ordering compared to sorting them by the number of correct ratings.
This suggests that participants who rated the veracity of less articles while taking more time judging each one achieved higher accuracy of their veracity ratings.
To further study the differences between participants who were successful and unsuccessful in rating the veracity of articles, we separated out two groups participants: \emph{successful group} consisting of 25\% of participants with the highest accuracy of their ratings, and \emph{unsuccessful group} consisting of 25\% of participants with the lowest accuracy of their ratings.

There is a significant difference (p=0.008 using unpaired t-test) in the distribution of average time spent looking at headings of posts in the feed in the evaluation phase of the experiment between the successful and unsuccessful groups.
The group of successful participants spent on average 2 seconds (SD 0.6) per post in the feed looking at its heading, while the group of less successful participants spent on average 2.9 seconds (SD 0.8).
There is not a significant difference in the time spent observing images or leading paragraph texts (perex) of posts in the feed.

\begin{figure}[t]%
    \centering
    \includegraphics[width=.8\linewidth]{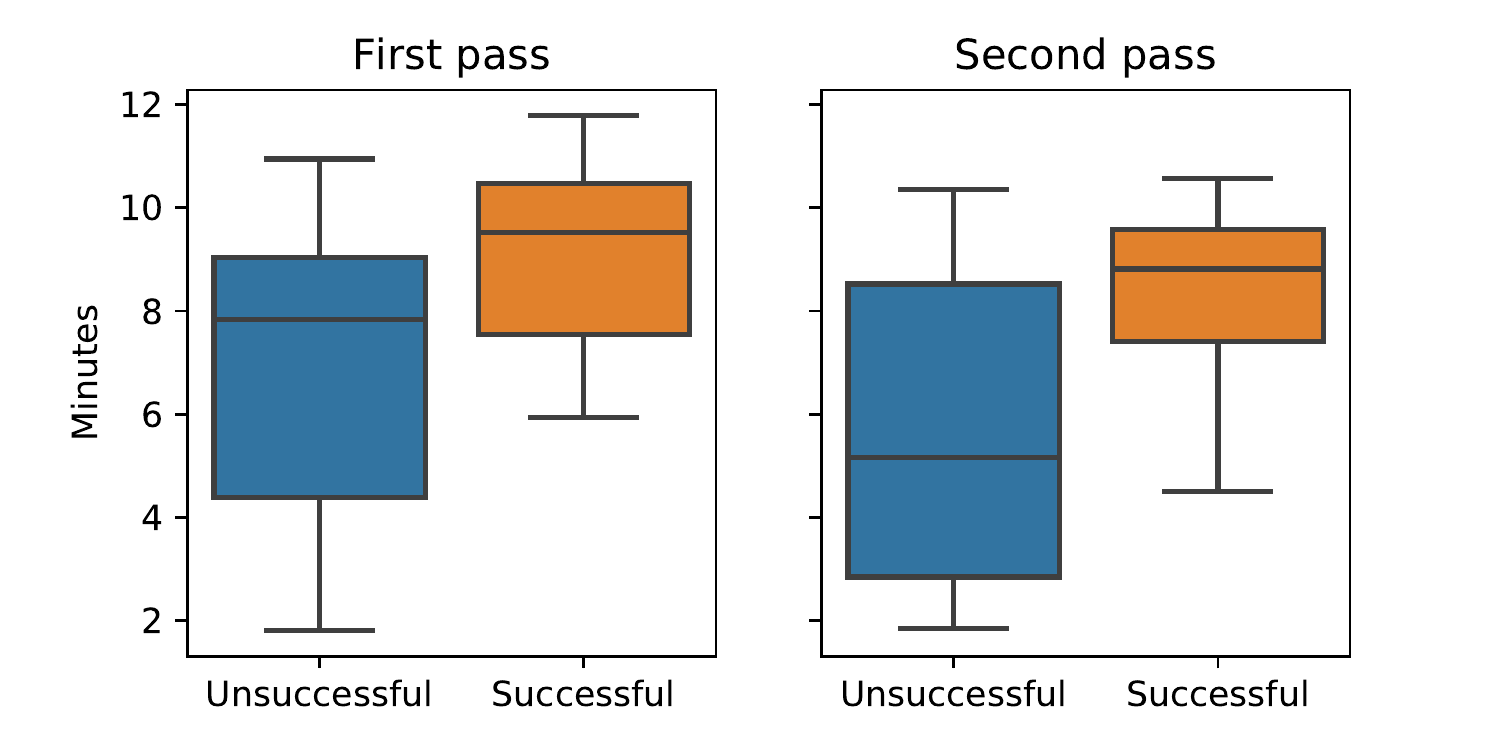}%
    \caption{Difference in total time spent reading articles between a group of 25\% most successful participants (successful participants) in rating the veracity of articles and 25\% least successful participants (unsuccessful participants).
    Successful participants spent on average 8 minutes (SD 3) reading articles in the first pass and 7 (SD 3) in the second pass.
    Unsuccessful participants spent on average 7 minutes (SD 3) reading in the first pass and 5 (SD 3) in the second pass.}%
    \label{fig:total-reading-time-successful-vs-unsuccessful}%
\end{figure}

\begin{figure}[t]%
    \centering
    \includegraphics[width=.8\linewidth]{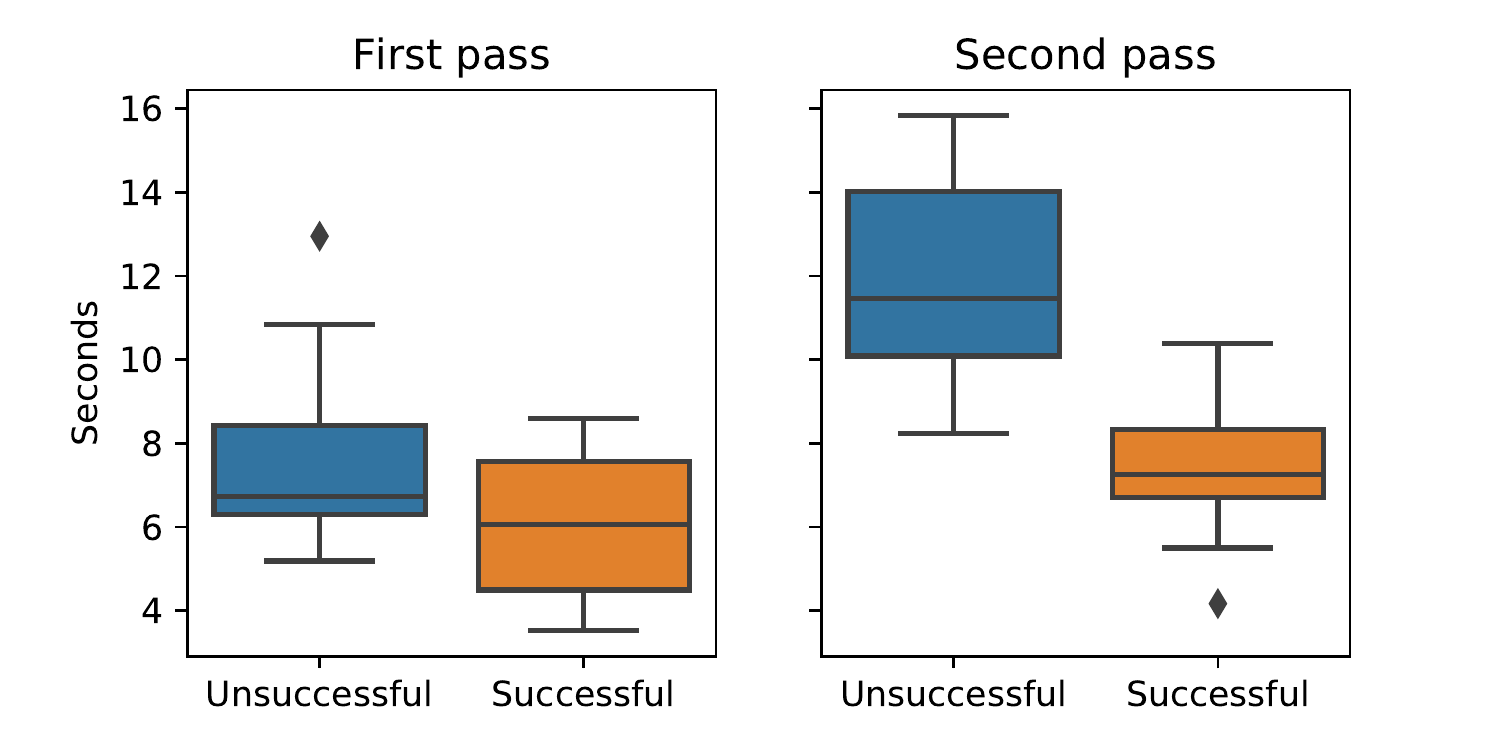}%
    \caption{Average time spent looking at posts in the feed per participant for the groups of top 25\% most and least successful participants. Unsuccessful participants relied on information given in the feed, which can be seen by the longer dwells on posts especially in the second pass of the experiment.}
    \label{fig:time-spent-looking-at-posts-succesfulness}%
\end{figure}

This suggests a higher reliance on information given in the feed for participants that were less successful compared to more successful participants.
The finding is supported by the comparison of total time spent reading articles in the first and second passes of the experiment shown in Figure~\ref{fig:total-reading-time-successful-vs-unsuccessful} as well as the comparison of time spent looking at posts in the feed shown in Figure~\ref{fig:time-spent-looking-at-posts-succesfulness}.

\subsection{Study limitations}

\begin{figure}[htb]%
    \centering
    \includegraphics[width=.7\linewidth]{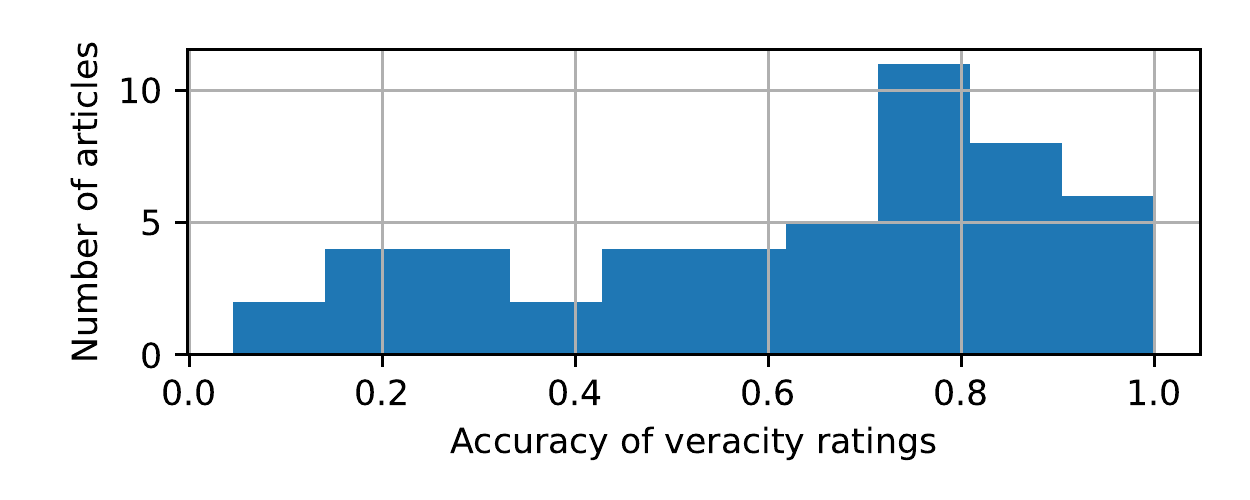}%
    \caption{Histogram of accuracy of veracity ratings of articles. Articles with accuracy of 1 received only correct veracity ratings. ``Couldn't tell'' ratings are not considered.}
    \label{fig:accuracy-articles-histogram}%
\end{figure}

As we proceeded with the data analysis, we noticed a variability in how difficult it was for the participants to evaluate the veracity of individual articles (i.e., \emph{article difficulty} for short). During the preparation of the stimuli, we did not control for the article difficulty. Despite most of the articles appear to be reasonably difficult to evaluate, the entire set of the articles ranged from ``almost impossible'' to ``very easy''. Figure~\ref{fig:accuracy-articles-histogram} illustrates the distribution of article difficulty using a histogram of accuracy of veracity ratings of articles.
While most articles (70\%) were better than random guessing would evaluate them, 8 articles (16\%) achieved accuracy of 0.25 or less, which means that they were not only hard to evaluate, but were outright deceitful. Such varying difficulty could potentially skew some future findings and in further similar studies and we suggest to control the difficulty through piloting.

Other limitations follow from the study being carried out in the controlled laboratory setting instead of a more natural setting of a real social network. In our study, we purposefully omitted the source of news as well as social media signals, such as likes, comments, etc. to reduce the number of (context) variables and focus on the content signals. 
Also, our sample of participants consisted solely of high school students (prospective university applicants); however, we did not control for the geographic location of their high school or its type, nor for the socioeconomic backgrounds of the participants All these might be factors influencing participants' opinions or their level of critical thinking. In general, the participants were lay persons in terms of news veracity evaluation. We felt, that we needed to run a similar study with fake news detection professionals, whose behavior can differ from the current group. Thus, we conducted a follow-up study, described in the next section.

\section{Follow-up study: qualitative with experts}

Given the results of the main study, we followed-up on them by conducting another experiment. It was based on the same general design with one key difference: \emph{we invited seasoned fake-news annotators (i.e. fact-checkers)} as participants. These \emph{experts}\footnote{We use the term ``expert'' in the sense of ``fact-checking experts'' not ``topic/domain experts''.} have the experience of recognizing fake news either as professionals or activists.

The main goal of the follow-up study was to further explore the behavior of successful news veracity evaluators. As the pool of experts that could participate was limited, only 7 participants managed to take part in the experiment. We needed to alter the original study design as follows:
\begin{itemize}
    \item \emph{The analysis was now predominantly qualitative.} The study now included a structured interview and a session retrospective in addition to the original scenario. In these additional steps, the experts explained their strategies and decisions in-depth.
    \item \emph{Each expert participated in a separate session} (i.e., there were no group sessions), which was guided by a study moderator.
    \item \emph{The first task from the main study (news consumption) was omitted}. This was due to the fact, that it required the participants to be oblivious to the purpose of the entire study. Such condition could hardly be satisfied when fake-news experts were invited. Also, this task is really more suited for the behavior examination of ``general social network population''. With fake-news experts, we were more interested in their behavior during the veracity evaluation task, which remained in the study scenario (along with the questionnaire).
    \item \emph{There was no time limit for the news veracity evaluation task.} As we had access to only limited number of expert participants, we removed the 12 minute time limit to record as much behavior as possible.
\end{itemize}
Other study aspects were preserved. We used the same stimuli (the same articles and ``fake Facebook'' application). We also recorded experts' gaze albeit with a different purpose: the recording was used immediately in the session to support the recollections of the participants during the retrospective.

\subsection{Research questions}
\label{rqas}

The follow-up study had the following research questions:

\begin{description}
    \item[RQ-Q1] What behavior characterizes the experts during news veracity evaluations? Was it different to main study participants (with respect to the study conditions)?
    \item[RQ-Q2] What features do experts use for their evaluations? In particular, we also examined the following sub-question:
    \begin{description}
        \item[RQ-Q2.1] What information do the experts miss for definitive verification? Particularly, what is the role of appearance of the news content, news source or other contextual metadata?
    \end{description}
    \item[RQ-Q3] What characterizes the articles that were more difficult to evaluate?
\end{description}

\subsection{Participants}

We invited journalists and/or media experts to participate in the follow-up experiment. First, we created a list of potential participants by querying contacts we had from the past to the Slovak journalist scene. From them, we gathered references to people that could potentially participate in the experiment: we targeted individuals, that had at least some experience with manual fake news detection. Then, we requested their participation via e-mail.

Our request was positively answered by 7 participants (denoted as E1 to E7). All of them have at least some work experience in the media sector. Some of them are journalists who regularly write about the fake news topic, others actively participate on debunking of fake news (e.g., as curators of fact-checking websites). The participants were all males in their late 20ties or early 30ties.

\subsection{Environment and equipment}

The follow-up study took place in the \emph{single-participant eye-tracking laboratory} of the User Experience and Interaction Research Center. This laboratory complements the group laboratory used for the main study and is better suited for the qualitative character of the follow-up study. It is placed in a small \emph{experiment} room, that houses one workstation mounted with Tobii TX300 eye-tracker (integrated with participant's screen) and one observation screen for the study moderator (who is seated across the table opposite of the participant). Adjacent to this room is a larger \emph{observation} room, that allows additional people (e.g., event annotators), to observe the experiment through a one-way mirror and through digitally transferred screen and voice.

The follow-up study used the same experimental application (the ``fake Facebook'') as was used in the main study. The screen resolution was kept to the original 1900x1200. The Google Forms questionnaire was re-used as well. The recording was done through Tobii Studio software, which also implements the \emph{retrospective} feature.

\subsection{Study scenario (tasks)}

The session followed this scenario:
\begin{description}
    \item[Introduction.] The moderator explained the purpose of the study to the participant (i.e., assessment of expert fake-news annotator behavior) and briefly introduced the concept of eye-tracking. Then, the moderator outlined the scenario of the study to the participant, putting focus on the explanation of the \emph{veracity evaluation task}. After that, the participant could ask questions about the study procedure. Lastly, the participant undertook the eye-tracker calibration.
    
    \item[Step 1: News veracity evaluation.] This activity was the same as in the main study: the particpant's task was to indicate, whether he considers the presented posts/articles as true or false. The participant could scroll through the feed, read the posts, open-up and read the full articles and indicate the veracity using buttons placed below each post. Unlike in the main study, the participant could take as much time as he needed for this step. On average, it took the participants 27 minutes to finish this task. \emph{We used the same stimuli (posts/articles) as in the main study}.
    
    \item[Step 2: Questionnaire.] We used the \emph{same questionnaire as in the main study}, mostly for control purposes. The low number of participants did not allow a reliable quantitative comparison to the lay participants of the main study.
    
    \item[Step 3: Structured interview.] In this step, the moderator asked the participant several questions (see below) regarding the techniques of fake news detection.
   
    \item[Step 4: Retrospective.] The recording of the news veracity evaluation task was re-played to the participant. The participant observed the screen recording overlaid with his own gaze. In a dialogue with the moderator, the participant commented the recording and answered couple more questions.
\end{description}

In the structured interview (step 3), the goal was to gather the explanations of the participant's actions as well as his views on the nature of the news veracity task. The interview was semi-structured: the goal of the moderator was to gather answers for all of the questions listed below, but there was no strict rule in what order the questions were to be asked. This was left to the discretion of the moderator based on the actual behavior and answers of the participant. However, the moderator made sure that he received comments helping to answer the research questions laid out in the  section~\ref{rqas}. The same principle was applied also to the retrospective (step 4), which somewhat complemented the structured interview.

The particular questions asked in the structured interview were the following:
\begin{itemize}
    \item How do you understand the concept of fake news?
    \item How do you evaluate fake news in your own practice?
    \item What information do (you think) you used for evaluation of news veracity in this experiment?
    \item What information did you miss?
    \item Would you evaluate the veracity of an article based also on the structure and appearance of the web page it is placed in?
    \item Are you able to evaluate the veracity of an article based solely on its metadata (i.e., without reading the text itself)?
    \item Is the difficulty of veracity evaluation varying between the tasks (in this experiment and in general)? What are the differences?
\end{itemize}

In the retrospective, further questions were asked (repeatedly):
\begin{itemize}
    \item Why did you decide to pay more/less attention to this particular article?
    \item Why did you decide to pay more/less attention to this particular detail?
\end{itemize}

\subsection{Process and analysis protocol}

The study was done in 7 sessions over several days. With each participant, we followed the given scenario. The entire sessions were recorded as audio, video (screen) and gaze. 

Afterwards, the recordings were analyzed. First, the participants' answers were transcribed. To organize the diverse and heterogeneous set of ``free-text'' answers, we performed a top-down analysis driven by research questions. The purpose was to seek participants' expressions relevant to particular research questions and quantify their occurrence across all participants. For this, we needed to establish a set of stereotypical answer codes and assign them to the original answers. 

The coding was done in two passes. In the first pass, the annotator reviewed all participants' answers and created the candidate answer codes (e.g., \emph{the participant expresses the importance of news source credibility}). Then, a reviewer reviewed the potential set of codes. In the following second pass, the original participants' answers were assigned to these codes. The assignment was then checked once more by the reviewer.

\begin{figure*}[t]%
    \centering
    \includegraphics[width=.7\linewidth]{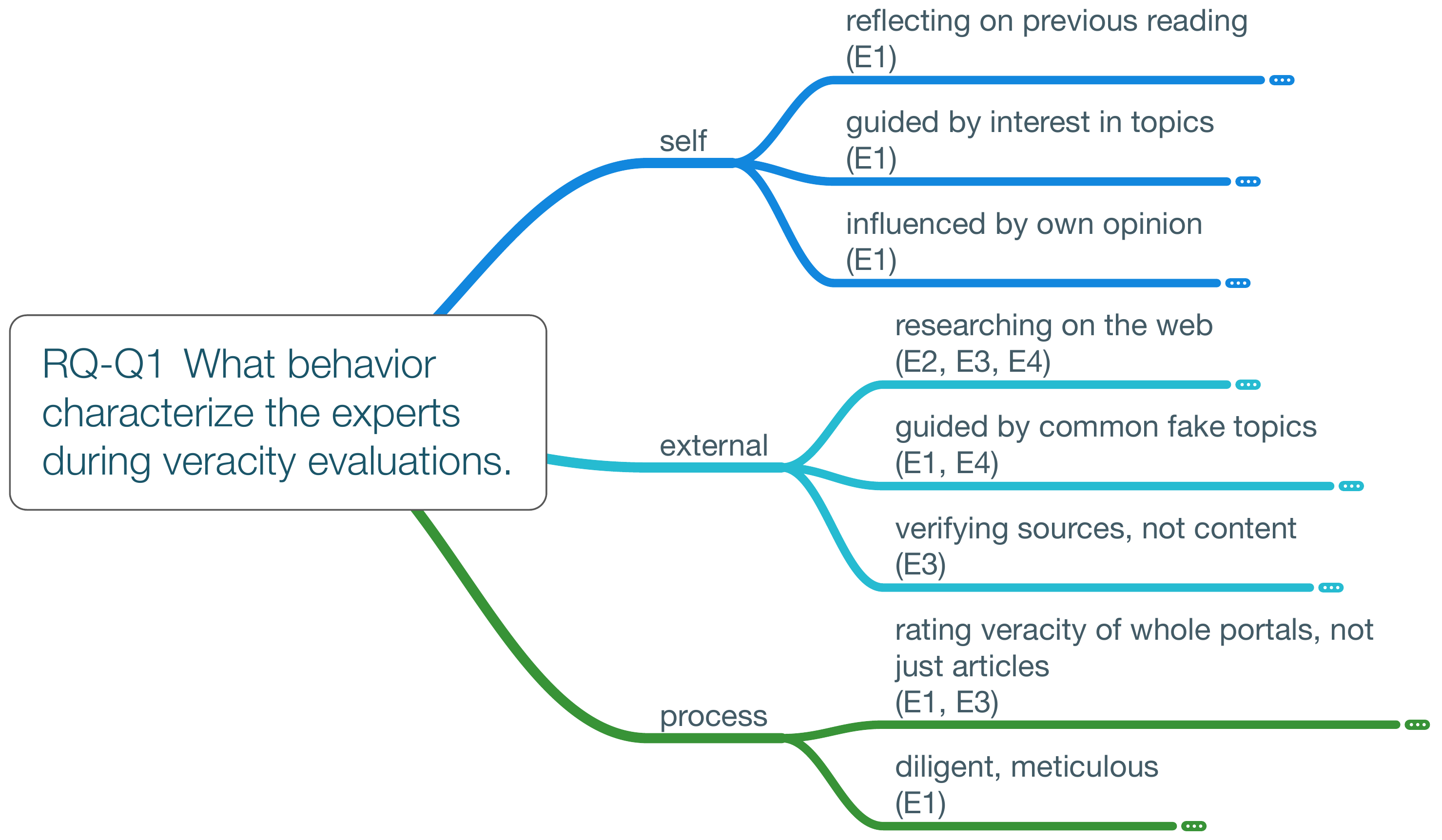}%
    \caption{
        Codes for RQ1.
    }%
    \label{fig:coding-rq1}%
\end{figure*}

\begin{figure*}[t]%
    \centering
    \includegraphics[width=.7\linewidth]{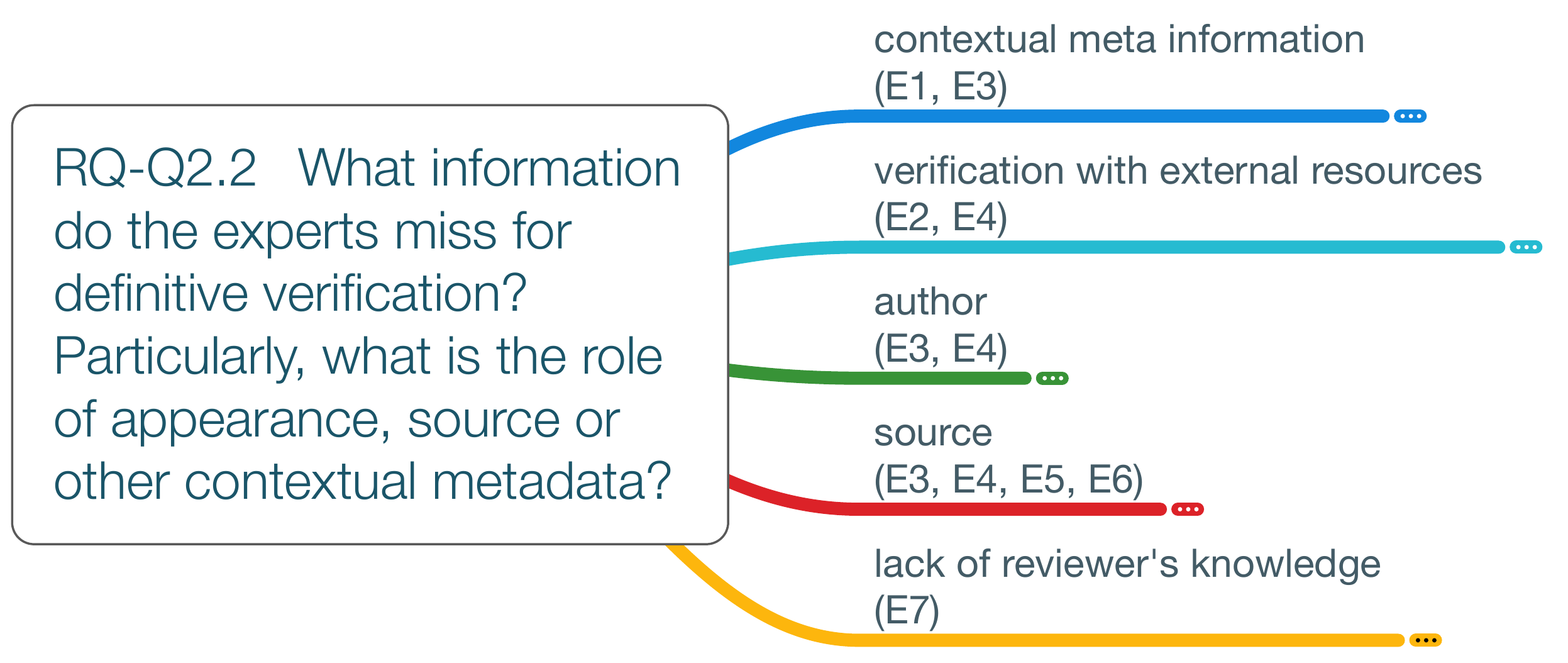}%
    \caption{
        Codes for RQ2.1.
    }%
    \label{fig:coding-rq21}%
\end{figure*}

The resulting codes and their occurrences can be viewed in diagrams in the figures~\ref{fig:coding-rq1},~\ref{fig:coding-rq2},~\ref{fig:coding-rq3}~and~\ref{fig:coding-rq21}.

The experiment also yielded other data, namely veracity evaluations. All 50 articles were evaluated by 5 experts. Expert E6 evaluated 48 articles and expert E1 evaluated 35 (in the beginning, we were not emphasising the instruction to evaluate all articles).

\subsection{Findings}

Based on the data collected and answers coded, we are able to formulate several findings of the follow-up study, which we organize by the given research questions.

\subsubsection{RQ1 What behavior characterizes the experts during veracity evaluations?}

% distribucie hodnoteni

\begin{figure*}[t]%
    \centering
    % \begin{mdframed}[linecolor=blue, linewidth=3pt] % TODO remove for CR
    \includegraphics[width=\linewidth]{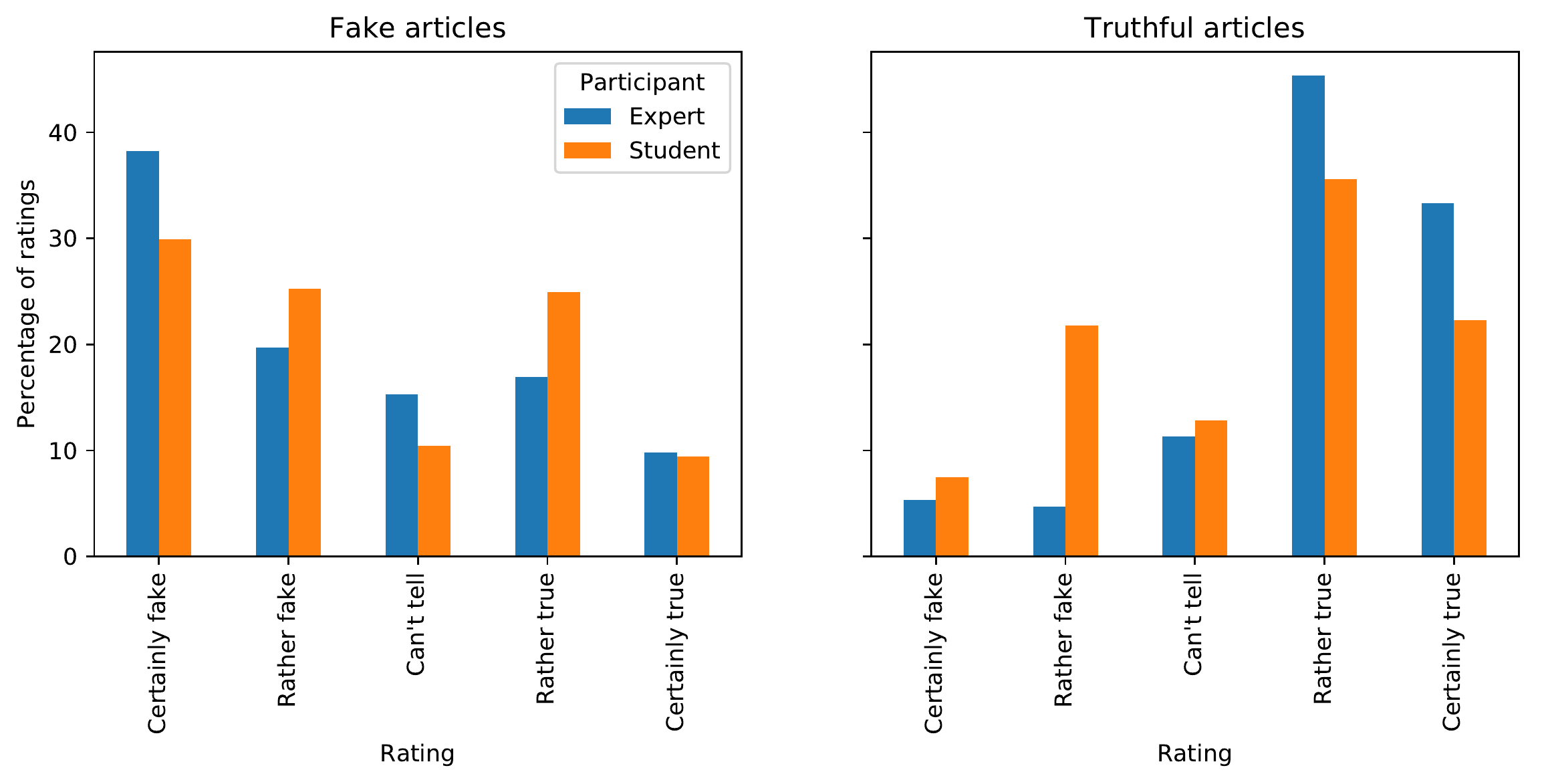}%
    \caption{
        Distribution of veracity ratings made by students and experts.
        Expert participants were more successful in their veracity ratings, especially for truthful articles.
    }%
    \label{fig:expert-vs-student-ratings}%
    %\end{mdframed} % TODO remove for CR
\end{figure*}

Figure~\ref{fig:expert-vs-student-ratings} shows the distribution of veracity ratings made by expert and student (main study) participants.
Overall, expert participants were more successful in their veracity ratings than student participants: overall accuracy of experts was 69\% while the accuracy of students in the main study was 56\%. The ``can't tell'' answers are counted as unsuccessful.

The better performance of experts could be expected because of their experience as well as no time constraint. A more interesting observation is, that in more than 30\% of cases, the experts were still unable to correctly guess the truthfulness of an article based solely on its content. This indicates, that additional metadata are indeed required for successful evaluation. This was also confirmed by the experts in the interviews.

A more surprising observation was, that the accuracy increase was not evenly distributed over all articles. Rather, a larger difference in accuracy was observed for truthful articles. There, 79\% of expert answers were correct (compared to 58\% accuracy of students). On the other hand, for fake articles, the experts were only 58\% accurate (compared to 55\% accuracy of students). 

This difference (20 percentage points versus 3 percentage points) is quite striking and we cannot offer any certain explanation for this. One possible explanation, which is in line with expert answers (listed further below), is that the truthful articles usually contain more factual statements that can be verified. If at the same time, the experts have knowledge about the related topics, which they reportedly use for their initial appraisals, then their judgement can be more accurate.

% TODO other observations from behavior data
% TODO findings from the answers

% porovnanie uspesnosti studentov vs expertov
% Z dôvodu nevhodných predpokladov na Studentov T-test, sme použili neparametrickú verziu Mann Whitney U. Na základe porovnania výsledkov Mann Whitney U sme zistili, že p=1.859e-07, čiže p<0.001.
% Z toho vyplýva, že pravdepodobnosť chyby 1. rádu je veľmi nízka a to naznačuje, že rozdiel v presnosti určenia medzi skúsenými a bežnými používateľmi bol štatisticky signiﬁkantný.

% porovnanie oblasti fixacii v clankoch – tam vyzera, ze sa spravaju aj experti aj studenti podobne a heatmapa sa velmi nelisi

% porovnanie casu straveneho citanim clanku – experti 18 voci studenti 7 minut. je to ale skreslene tym, ze studenti mali obmedzeny cas a experti nie
% no time restriction (took about 27 minutes, 18 minutes reading articles)

% porovnanie uspesnosti v temach so silnym/slabym postojom pre expertov a studentov

% dalsie napady, ktore nemame vyhodnotene:
% - rozdiel medzi hodnoteniami studentov a expertov – ci ta sila presvedcenia hodnotenia “urcite” alebo “skor pravdivy/nepravdivy” nejako inak vplyvala na ich uspesnost u oboch skupin, ale mozno to uz je trocha mimo tej nasej vyskumnej otazky..
% - porovnanie rychlosti hodnotenia studentov a expertov

\subsubsection{RQ2 What features do experts use for their evaluations?}

\begin{figure*}[t]%
    \centering
    \includegraphics[width=\linewidth]{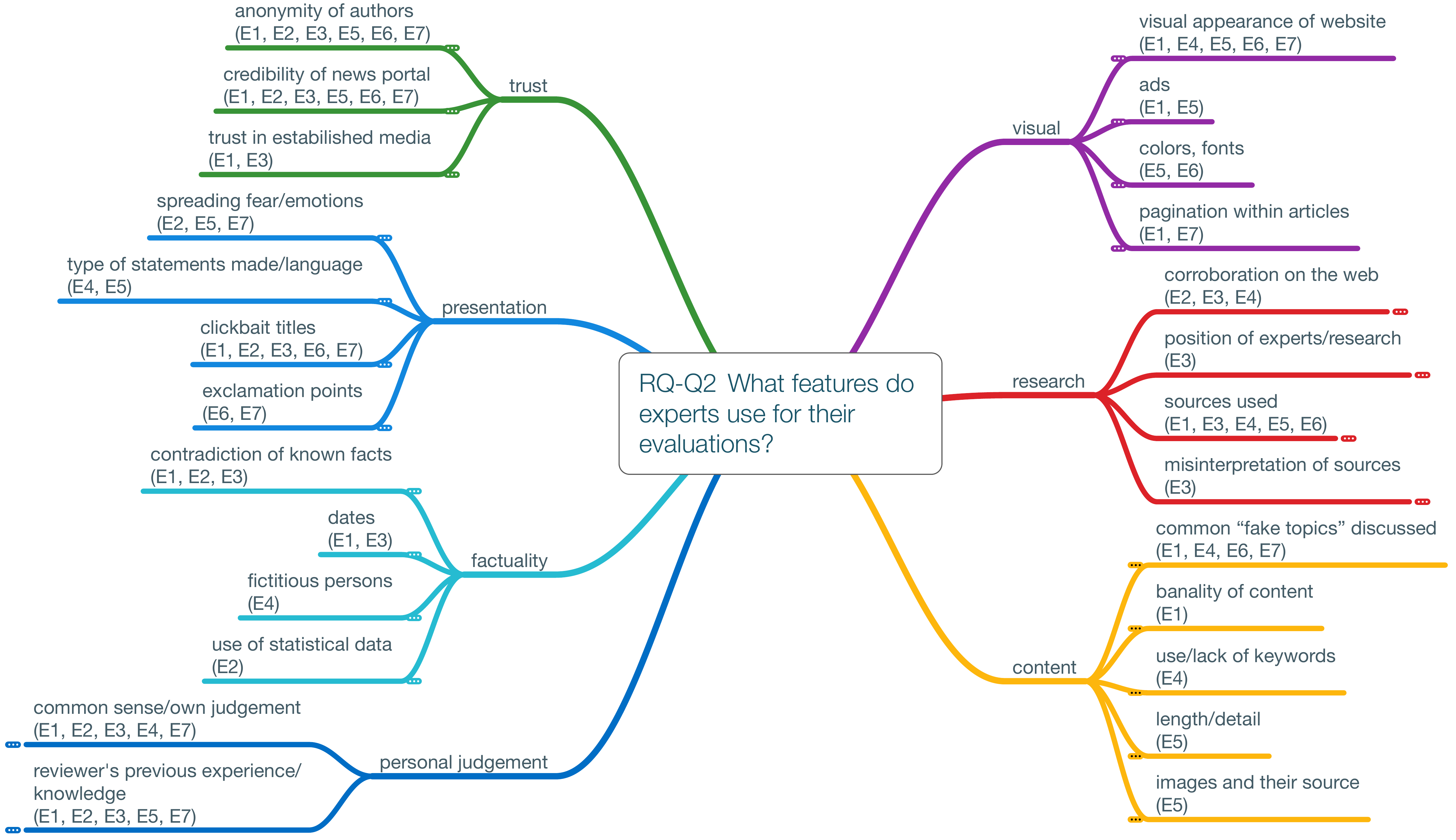}%
    \caption{
        Codes for RQ2.
    }%
    \label{fig:coding-rq2}%
\end{figure*}

Figure~\ref{fig:coding-rq2} gives an overview of features the experts reflected on.
This section explores the features in more detail.

\paragraph*{Trust and credibility}
A common and repeating point among almost all reviewers was the determination of trust and credibility of authors and publishers of judged content.
Anonymity of authors was a strong indicator of fake content for the reviewers.
Furthermore, the reviewers tended to be skeptical of media portals that were unknown to them.
On the other hand, well-known and established authors as well as media portals increased the trust of reviewers in the content.
Several reviewers also mentioned well-known fake news portals which content they were likely to quickly discard as fake.
Expert E5 summarized his position as follows:
\emph{``In good and bad, there are strong brands and names of journalists, which when spotted, I am certain the content is not fake and, reversely, there are persons and web portals that are so untrustworthy that I know their content is fake.''}

\paragraph*{Presentation of content}
The way that articles present information was indicative of their trustworthiness for the experts.
According to the experts, language of fake content commonly includes the use of labels, judgemental statements and comments.
Articles that aim to build up emotions such as fear in readers were a sign of fake content for experts E2, E5, and E7.
The experts mentioned the use of clickbait titles and exclamation marks as common in such articles.

\paragraph*{Use of facts}
The experts looked for contradictions of known and verified facts when fact-checking articles.
They reflected that they specifically look for factual information such as dates of events and statistical data.
When stated, the facts can be verified by reviewers.
However, also the lack of factual information was indicative of fake content for the experts.
They also mentioned their effort to identify fictitious persons.

\paragraph*{Personal judgement of reviewers}
The reviewers reflected that they often rely on the common sense and their own judgement during reviews.
Expert E7 responded that he employs
\emph{``sane reasoning, if a nonsensical article is being spread, we can evaluate it based on knowledge from different fields.''}
Reviewers further said that they evaluate whether an information is in line with their existing knowledge:
\emph{``If something matches my thought pattern, I make a decision.''} (E1)
They evaluate trivial articles in a similar fashion:
\emph{``It seemed banal, it couldn't be untrue.''} (E1)

\paragraph*{Visual appearance of websites}
The reviewers mentioned several visual features that are common for fake web articles.
They are surrounded by lots of ads.
They make use of radiant colors, large and bold fonts.
They often contain pagination within content.

\paragraph*{Research}
Researching the content of articles was a common task for the reviewers.
They look for corroboration of information by searching on the Web and within news aggregators (e.g., Google News).
They look for positions of experts, researchers and verified sources.
Furthermore, they examine the sources referenced within articles to check for their trustworthiness and potential misinterpretation of them.

\paragraph*{Content}
The reviewers often referred to common ``fake topics'' that are primarily discussed in fake articles.
Expert E1 said:
\emph{``There are evidently fake articles that are repeated over and over again: vaccination, anti-vax movement, immigrants are threatening us, LGBTI disease.''}
Therefore, topics of content discussed may be used as features.
The reviewers mentioned several other features relating to the content of articles.
Such are the length of articles and their level of detail.
Use (or lack of use) of certain keywords may be suggestive of fake content.
Finally, images used in articles can be verified whether they belong to the described topic and for their use elsewhere on the Web.

\subsubsection{RQ2.1 What information do the experts miss for definitive verification?}
% Particularly, what is the role of appearance, source or other contextual metadata?

Reflecting on the user study setup, the reviewers said that they mainly missed contextual information such as authors and sources of articles.
Furthermore, they missed the ability to search for and verify information on the Web which is common practice for them when reviewing.
Expert E2 stated that: \emph{``What I mind about this research is that usually I tend to Google and look for official sources. I try to factually verify it.''}
In effect, they lacked knowledge from the reviewed topics that would let them make informed decisions.

\subsubsection{RQ3 What characterizes the articles that were more difficult to evaluate?}

\begin{figure*}[t]%
    \centering
    \includegraphics[width=.7\linewidth]{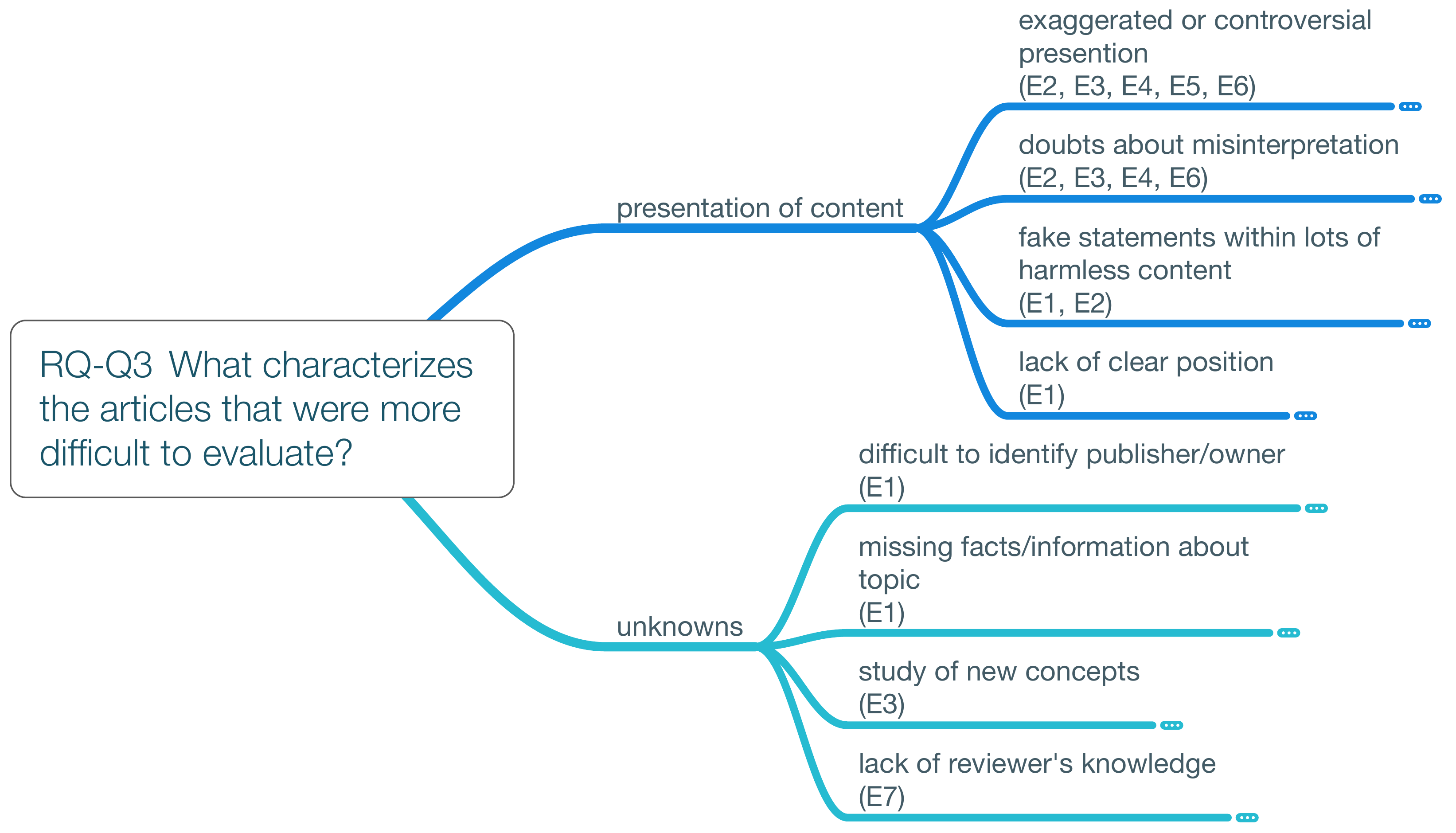}%
    \caption{
        Codes for RQ3.
    }%
    \label{fig:coding-rq3}%
\end{figure*}

Codes identified within this research question are presented in Figure~\ref{fig:coding-rq3}.
We grouped them into two categories: presentation of content and unknowns.

\paragraph*{Presentation of content}

A repeating characterization of articles that were difficult to evaluate by the reviewers was their misleading presentation of potentially factual information.
Factual information may be presented in a controversial light, or exaggerated such that it is no longer reliable.
Expert E4 remarked that \emph{``...although the factual information may be truthful, they are presented in a dismissive context or commented in a misleading way with the goal of manipulation or pure lie.''}
Such articles may also misinterpret their sources or combine truthful information from various sources but connect it misleadingly.
As difficult, the experts also labelled the articles that bury small amounts of misinformation among lots of harmless content, for instance some anti-vaccination remarks within a portal containing recipes for parents (E1).
Finally, articles that only suggest a misleading stance but lack a clear position and suggest the user to make up their own mind are also difficult to judge.

\paragraph*{Unknowns}
The reviewers suggested that articles may be difficult to evaluate because of a lack of certain information.
Expert E1 remarked that it may be difficult to identify publishers or owners of Web portals publishing fake content.
Articles may also miss significant information or facts about the discussed topics such as dates of events that would enable the reviewers verify them.
Finally, such articles may require the reviewers to study concepts that are new to them or go beyond their knowledge from discussed fields or topics.

% distribution of veracity ratings of experts vs students/opinion strengths for topics of experts vs students
\section{Conclusion and future work}

We have presented two studies (main and follow-up) where behavior of human participants was recorded during their interaction with (fake) news in a social media interface. The concept of the main study answers the demand expressed by related literature: the research of the characterization and detection of fake news can benefit from human behavior analysis. This stands (1) for the news consumption by ``unsuspecting'' users and (2) for the news veracity evaluation, when users actively look for veracity clues. Observation of the former serves the general understanding of the fake news phenomenon. The latter can guide the creation of automated detection approaches.

Our contributions are the (1) design of the studies, (2) the resulting dataset of the main study and (3) findings that result from the data analysis. The main study findings include:
\begin{itemize}
    \item Reading behavior of participants was less influenced by their interests when tasked to rate veracity of articles compared to browsing of the feed without this goal in mind.
    \item Explicit actions (e.g, likes, shares) were seldom used spontaneously, which was expected because of the limitations of the study. Their indicative power (towards fake detection), thus, remains questionable. The exception is the \emph{report} action. Out of 12 times it was used, 10 times it pointed to fake articles.
    \item In contrast with participants that were successful in estimations of the veracity of articles, unsuccessful participants relied more heavily on information given in the feed as opposed to actual full articles.
    \item Strength of opinion reflects the strength of interest of participants for most of the topics. However, in several topics where the interest was generally lower, the participants had relatively strong opinions with a common polarity. This is hard to explain, but it may suggest that people are often affected by external environment to form their opinions.
\end{itemize}

After the main study was evaluated, we saw an opportunity to conduct a similar study with professional fact-checkers (dubbed the ``experts'') as participants. This follow-up study yielded further observations, most important of which are the following:
\begin{itemize}
    \item Experts were more successful in veracity evaluation as expected, but surprisingly the increase was almost entirely attributed to truthful articles.
    \item Even experts cannot solely rely on the content of the articles, but require additional metadata about them.
    \item From all possible content and context features of the news, the expert annotators  prize information on article source (i.e. author and website identity). They also prize the presence of labels, judgemental statements and comments, which are characteristic for fake news, and also the presence of factual statements, which can be verified. They also give importance to the topics featured in the articles: the experts often referred to certain topics that are typical for fake news. Further features mentioned by experts included visual appearance, formatting or length of the articles as well as excessive presence of ads. %With exception of the article source features, this list
    \item Experts further pointed out that they rely on their own knowledge when appraising the truthfulness and ultimately, on additional research and cross-checking of the facts with other credible sources. As a particular practice, the experts mentioned the use of image search for verification of images adherence to the presented textual content. 
\end{itemize}

% TODO skomentovat pouzitelnost ficur pri automatickej detekcii.
%This implies, that a truly successful automated fake news detection approach must somehow incorporate wide spectrum of existing knowledge.

We see multiple avenues for future work. The study concept can (and should) be deployed for more demographic groups. Once the pool of recorded data is larger, more observable stable behavior patterns will emerge. This will allow the introduction of more independent variables into the design. Namely, more information features can be added to the current (rather reductionist) user interface. In the current setting, social interaction elements such as comments, likes, shares etc. are missing and it is relevant to ask how (if at all) the users take them into account in both tasks. Also, the heterogeneity of article sources and visual design (which was withheld in our current work) can be thrown into the play.
%OPT some literature hints that the social interaction indicators do have the impact on the likeability of post sharing (niekde som to cital, ale neviem uz kde, bolo by treba znovu vyhladat)

\section*{Acknowledgements}

This research was partially supported by TAILOR, a project funded by EU Horizon 2020 research and innovation programme under GA No 952215.

\bibliographystyle{apacite}
\bibliography{references}

\end{document}